\documentclass[sigconf]{acmart}

\setcopyright{none} 
\acmConference[Accepted to ACM CCS 2022]{ACM Conference on Computer and Communications Security}{7-11 November 2022}{Los Angeles}
\acmYear{2022}

\usepackage{multirow}
\usepackage{appendix}
\usepackage{hyperref}
\usepackage{tabularx}
\usepackage{booktabs}
\usepackage[font=small]{caption}
\usepackage{subcaption}
\usepackage{units}
\usepackage{xspace}
\usepackage{enumitem}
\usepackage{rotating}
\usepackage{todonotes}

\usepackage{hyperref}

\newcommand*{\eg}{e.g.,\@\xspace}
\newcommand*{\ie}{i.e.,\@\xspace}
\newcommand{\enquote}[1]{``#1''}

\makeatletter
\patchcmd{\hyper@makecurrent}{%
    \ifx\Hy@param\Hy@chapterstring
        \let\Hy@param\Hy@chapapp
    \fi
}{%
    \iftoggle{inappendix}{
        \@checkappendixparam{chapter}%
        \@checkappendixparam{section}%
        \@checkappendixparam{subsection}%
        \@checkappendixparam{subsubsection}%
        \@checkappendixparam{paragraph}%
        \@checkappendixparam{subparagraph}%
    }{}%
}{}{\errmessage{failed to patch}}

\newcommand*{\@checkappendixparam}[1]{%
    \def\@checkappendixparamtmp{#1}%
    \ifx\Hy@param\@checkappendixparamtmp
        \let\Hy@param\Hy@appendixstring
    \fi
}
\makeatletter
\newtoggle{inappendix}
\togglefalse{inappendix}

\apptocmd{\appendix}{\toggletrue{inappendix}}{}{\errmessage{failed to patch}}
\apptocmd{\subappendices}{\toggletrue{inappendix}}{}{\errmessage{failed to patch}}

\usepackage[utf8]{inputenc}
\usepackage[T1]{fontenc}
\DeclareUnicodeCharacter{03B5}{\ensuremath{\varepsilon}}

\begin{document}

\title[\enquote{Am I Private and If So, How Many?}]{\enquote{Am I Private and If So, how Many?} ---  Communicating Privacy Guarantees of Differential Privacy with Risk Communication Formats}



\author{Daniel Franzen}
\email{daniel.franzen@fu-berlin.de}
\affiliation{%
  \institution{Freie Universität Berlin}
  \country{Germany}
}

\author{Saskia Nu\~nez von Voigt}
\email{saskia.nunezvonvoigt@tu-berlin.de}
\affiliation{%
  \institution{Technische Universität Berlin}
  \country{Germany}
}

\author{Peter Sörries}
\email{peter.soerries@fu-berlin.de}
\affiliation{%
  \institution{Freie Universität Berlin}
  \country{Germany}
}

\author{Florian Tschorsch}
\email{florian.tschorsch@tu-berlin.de}
\affiliation{%
  \institution{Technische Universität Berlin}
  \country{Germany}
}

\author{Claudia Müller-Birn}
\email{clmb@inf.fu-berlin.de}
\affiliation{%
  \institution{Freie Universität Berlin}
  \country{Germany}
}

\renewcommand{\shortauthors}{Franzen et al.}


\begin{abstract}
Every day, we have to decide multiple times, whether and how much personal data we allow to be collected. This decision is not trivial, since there are many legitimate and important purposes for data collection, for examples, the analysis of mobility data to improve urban traffic and transportation. However, often the collected data can reveal sensitive information about individuals. Recently visited locations can, for example, reveal information about political or religious views or even about an individual's health.
Privacy-preserving technologies, such as differential privacy (DP), can be employed to protect the privacy of individuals and, furthermore, provide mathematically sound guarantees on the maximum privacy risk. However, they can only support informed privacy decisions, if individuals understand the provided privacy guarantees.
This article proposes a novel approach for communicating privacy guarantees to support individuals in their privacy decisions when sharing data. For this, we adopt risk communication formats from the medical domain in conjunction with a model for privacy guarantees of DP to create quantitative privacy risk notifications.

We conducted a crowd-sourced study with 343 participants to evaluate how well our notifications conveyed the privacy risk information and how confident participants were about their own understanding of the privacy risk.

Our findings suggest that these new notifications can communicate the objective information similarly well to currently used qualitative notifications, but left individuals less confident in their understanding.
We also discovered that several of our notifications and the currently used qualitative notification disadvantage individuals with low numeracy: these individuals appear overconfident compared to their actual understanding of the associated privacy risks and are, therefore, less likely to seek the needed additional information before an informed decision.

The promising results allow for multiple directions in future research, for example, adding visual aids or tailoring privacy risk communication to characteristics of the individuals.
\end{abstract}



\maketitle

\section{Introduction}

We generate a large amount of mobility data daily, for example, when using online map services for navigation or when purchasing public transport tickets. Due to the ubiquitousness of smartphones, collecting location and mobility data has become a quasi-standard also for many other applications, and prediction algorithms use the data collected for various purposes (\eg~\cite{gkoulalas-divanis_privacy_2011,papadimitriou_mining_2015,alghamdi_crowd_2019}).
Mobility data is also essential for the development of urban areas~\cite{sostaric_data-driven_2021, sun_leveraging_2017}, for example, to identify where public transport can be improved, bicycle lanes can be added or how new pedestrian zones might influence a neighborhood. These tasks are vital also to meet current environmental challenges~\cite{sostaric_data-driven_2021, paffumi_european-wide_2018}. To obtain the necessary data for such legitimate tasks we have to rely on either data donations by individuals~\cite{jokinen_would_2021} or on existing data collections by third parties.
However, laypeople are often unaware~\cite{almuhimedi_your_2015,shklovski_leakiness_2014,balebako_little_2013}, but location data can reveal sensitive information~\cite{krumm_survey_2009}. 
They can be used to identify locations of interest (\eg home address, religious or political organizations), reveal daily routines (\eg doing exercises), show social relationships (\eg collecting a child from daycare, dating) and might also disclose certain health conditions (\eg repeated visits to a specific medical facility)~\cite{wernke_classification_2014}.

In order to protect individuals' privacy privacy-preserving technologies (PPTs) like \emph{Differential Privacy} (DP) have been developed. The concept of DP, originally proposed by Dwork~\cite{dwork_differential_2006}, refers to the idea that the output of a computation should not reveal any new information (or at least verifiably little) about an individual, while preserving the conclusion obtained from the whole data set. 
A central point in the definition of DP is the parameter $\varepsilon$ which acts as a \emph{privacy budget}. It limits the amount of information obtained about any individual. Based on the value chosen for $\varepsilon$ a mathematically sound \emph{privacy guarantee} can be computed, which describes an upper bound for the privacy risk of any individual~\cite{domingo-ferrer_limits_2021}. Since the privacy protection offered by DP depends heavily on $\varepsilon$, the chosen value is an essential piece of information for individuals to decide about sharing data using DP.

Several techniques have been developed (\eg PATE for machine learning~\cite{papernot_semi-supervised_2016}) over the last years that realize the idea of DP by adding carefully constrained noise to the results of computations~\cite{agrawal_exploring_2021}.
Companies, such as Apple and Google, or governmental agencies, such as the US Census Bureau, already employ DP (cf.~\cite{domingo-ferrer_limits_2021}). However, so far, individuals have been made aware of the use of DP by explaining some aspects of its functionality. Information about the chosen privacy budget $\varepsilon$ and the resulting quantitative guarantees are not communicated~\cite{cummings_i_2021}. Thus, in the current state, individuals cannot evaluate the privacy protection offered by DP and, therefore, cannot make an informed decision on sharing their data. Instead we need methods to communicate the quantitative privacy guarantees to individuals. However, as privacy risk, aside from DP, is hard to quantify, these methods have not been researched very well so far.

In our research, we take the first step to close this gap. We propose using \emph{quantitative privacy risk notifications} to convey the DP privacy guarantee, \ie the remaining privacy risk of sensitive information being exposed when sharing data. As opposed to existing research that explains the functionality of DP (\eg~\cite{xiong_effect_2020}), we adapt empirically validated research from the medical domain, where quantitative risk communication is an essential part of education and decision-support processes. In particular, we identify recommended medical risk communication formats, such as percentages (52\%) or simple frequencies (26 out of 50), which can also be used to communicate privacy risk.
In a crowd-sourced study, we compare the resulting quantitative privacy risk notifications to conventional qualitative DP notifications and investigate the effect of these notifications on individuals' understanding of privacy risks.

Utilizing our research, we want to enable people to become aware of the level of protection offered by DP. We assume that only then individuals can make an informed decision by weighing up the purpose of the data collection with the remaining privacy risk before sharing their personal data. With this research we want to make the following contributions:
\setlist[itemize]{leftmargin=5.5mm}
\begin{itemize}
    \item We propose the usage of quantitative privacy risk notifications to communicate the privacy guarantees of DP in an understandable way.
    \item We derive suitable, empirically validated risk communication formats from existing medical research and combine them with privacy risk values derived from a mathematical model of DP \cite{mehner_towards_2021} into quantitative privacy risk notifications. These notifications represent the  real-world privacy guarantees by DP accurately.
    \item We evaluate the understandability of quantitative privacy risk notifications incorporating different risk communication formats in a crowd-sourced user study and find that they perform equally well to currently used qualitative notifications.
    \item Based on our results, we discuss opportunities for future research on supporting informed decisions with quantitative privacy risk notifications.
\end{itemize}

The remainder of this paper is structured as follows: We provide a brief overview of the current state of DP in \autoref{sec:stateOfTheArt} and argue that a new communication concept for DP is necessary. We substantiate this by discussing recent research into the communication of DP and survey suitable formats from medical risk communication.
\autoref{sec:design} details how we design novel privacy risk notifications utilizing selected risk communication formats and 
in \autoref{sec:studyDesign}, we describe the design of the study to evaluate their understandability.
The results are presented in \autoref{sec:results} and discussed in 
\autoref{sec:discussion}. We conclude in 
\autoref{sec:futureWork} with limitations and future work.


\section{Privacy Risk - Challenges and Opportunities}
\label{sec:stateOfTheArt}

Our research is situated in the following context: An individual (\eg the user of an app) has been using a mobility service (\eg a public transport app) for some time. This mobility service collects personal data (\eg mobility data, such as routes traveled or choice of transport mode). One day, the mobility service provider requests the individual's consent of sharing the collected data with a third party (\eg public administration). For the individual, sharing the data involves a certain \textbf{privacy risk}, the risk that sensitive information (\eg home address) can be learned, for example, by combining data from different sources~\cite{sweeney_simple_2000}. 
For making an informed decision, the individual needs to understand this privacy risk. Since the data are being anonymized by using DP, the mobility service provider can computer a \textbf{privacy guarantee}. This guarantee provides an upper bound on the probability, that sensitive data can be learned about the individual.
With this information, the mobility service provider shows a \textbf{privacy risk notification}, a compact user interface containing easy to understand information on the privacy of the individual. Within this notification, the mobility service provider can present the privacy guarantees in different \textbf{risk communication format}, for example, \enquote{Your privacy risk is 52\%} or \enquote{Your data might be leaked in 26 out of 50 cases}.

The described context highlights that well-communicated privacy guarantees provided by DP can allow individuals to make informed privacy-preserving decisions. However, a challenge in this context is, how we need to design such privacy risk notifications. The communicated privacy risk should be derived from the defined privacy guarantee provided by DP. Thus, we provide in~\autoref{sec:defDP} a short introduction to DP and explain how privacy guarantees are provided. 
Furthermore, we summarize research on existing approaches on communicating DP to individuals (cf.~\autoref{sec:comDP}), which so far have focused on qualitative descriptions only. We, therefore, extend our focus into the medical domain, where quantitative risk communication has been studied over the last twenty years (cf.~\autoref{sec:communicatingrisk}).

In the following, we use the term \emph{individual} (\eg for the user of the mobility service) to emphasize that the collected data contain personal and potentially sensitive information about people.
We assume that these individuals are laypeople who have a limited knowledge of privacy measures and technologies used, and might have different levels of experience with statistics and risk representations. Accordingly, privacy risks must be communicated to these individuals with particular care and circumspection.
Additionally, we consider two groups of stakeholders, namely, \emph{data owners} and \emph{data consumers}.
\emph{Data owners} are companies or service providers that collect individuals' data (\eg the mobility service provider).
\emph{Data consumers} are public institutions, companies or other third
parties using this data for statistical analysis (\eg the public administration).

\subsection{Differential Privacy}
\label{sec:defDP}

Differential privacy is a mathematical property that aims to protect the privacy of individuals against the data consumer when querying information from a data set stored in a statistical database~\cite{dwork_differential_2006}.

The underlying principle of DP is to limit the impact of a single individual on the analysis outcome. More specifically, the presence or absence of an individual's data \emph{must not} lead to a significantly different outcome of the analysis. Formally, assume two data sets~$D_1$ and $D_2$ differing in exactly one entry. A function~$f$ provides $\varepsilon$-DP if for all such~$D_1$ and $D_2$, all outcomes~$S\subseteq \text{Range}(f)$ satisfy
\begin{align*}
	P[f(D_1) \in S] \,\leq\, \mathrm{e}^{\varepsilon} \cdot P[f(D_2) \in 
	S]\text{.}
\end{align*}

\begin{figure*}
    \centering
    \includegraphics[width=0.9\textwidth]{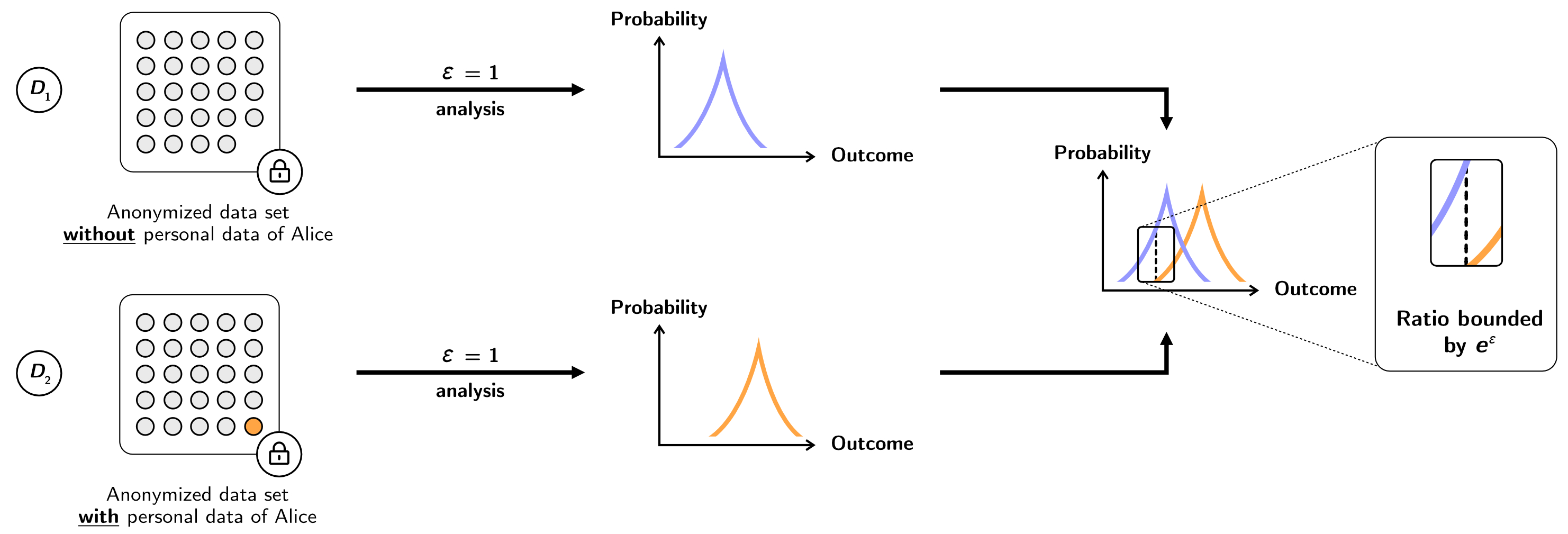}
    \Description[Dataflow of DP visualized with two dataset that only differ in one participant]{Two data sets are visualized which only differ in Alice's participation. DP calculates probability distributions which is close to the actual values. By the DP property the ration between the two distributions is bounded. With a certain probability the output could have come from either of the two data sets}
    \vspace{-1em}
    \caption{Visualization of the DP mechanism adding noise to two data sets.}
    \label{fig:dfn_dp}
    \vspace{-1em}
\end{figure*}

The core component of DP is the \textit{privacy budget} $\varepsilon$ that bounds how far an outcome of a data set is allowed to deviate when adding or removing one single individual.

We illustrate this property in \autoref{fig:dfn_dp} and explain it in more detail next.
We consider two data sets $D_1$ and $D_2$ that differ only in whether Alice has contributed her data (the red circle) or not.
Using DP to protect Alice, it is required that the resulting outcome of an analysis should be approximately the same independently of Alice's contribution.
To achieve this, the DP mechanism adds carefully tuned random noise to the outcome,
which results in a randomized outcome that is close to the actual value without DP~\cite{wood_differential_2018}.
In \autoref{fig:dfn_dp}, the distributions (red: with Alice, blue: without Alice) represent the probability of a certain outcome of the statistics.
The factor $e^{\varepsilon}$ bounds the ratio between these probability distributions, \ie how close the randomized outcomes of $D_1$ and $D_2$ have to be and, thus, determines
the maximum influence of a single individual's data point on the outcome. 
Since the probabilities of $D_1$ and $D_2$ are similar (only differing by the ratio that is set by $\varepsilon$), the data consumer cannot be certain whether $D_1$ or $D_2$ was used to generate the final outcome and, therefore, whether Alice is included in the data set or not.

The trade-off between privacy protection and accuracy of the result can be illustrated by the two edge cases: 
for $\varepsilon=0$, the definition requires the outcome distribution to be identical for both data sets ($P[f(D_1)] \,=\, P[f(D_2)])$.
Therefore, Alice's data does not influence the outcome. 
DP requires this property not just for Alice but for any two data sets that differ by one entry. By iteratively applying the same argument for other individuals, any results obtained could also be obtained from the empty data set.
Although this leads to perfect privacy, the outcome of $f$ is completely useless for any statistic purpose~\cite{wood_differential_2018}.
However, even with the complete randomization due to $\varepsilon=0$, a data consumer can still randomly guess whether Alice contributed towards the original result or not, and, thus, a trivial re-identification risk remains.

At the other extreme, \ie if $\varepsilon$ is very large, DP is already achieved with very little noise. The outcome of DP is very close to the actual value, almost as if no DP was used.
In this case, DP provides a high accuracy but low privacy.

In order to balance between the accuracy of the outcome and privacy protection, we need to choose a reasonable small $\varepsilon > 0$ that results in an outcome of $f$ which is \enquote{almost} independent of each single individual, at the same time, still accurately representing the body of the data as a whole. With such a choice for $\varepsilon$, DP essentially \enquote{hides} the data of each individual and its influence on the outcome.

When we carefully choose a value for $\varepsilon$, DP can be a valuable tool for protecting individuals' privacy.
By contrast, with a less careful choice of $\varepsilon$, the effectiveness of the privacy protection offered by DP can be reduced or even completely suppressed.
For this reason, the value used for $\varepsilon$ and the privacy guarantee resulting from it is essential information for evaluating the privacy protection offered by DP. Thus, providing this information to individuals is essential for making privacy decisions. In the next section, we give an overview of existing approaches to communicate DP.


\subsection{Privacy Risk Communication}
\label{sec:comDP}


Privacy decisions have become ubiquitous, not only due to legal requirements (\eg General Data Protection Regulation~\cite{the_european_union_regulation_2016}).
Their intention is to empower individuals during privacy decisions by requiring that data owners \enquote{shall implement appropriate technical and organizational measures} as part of \enquote{data protection by design and by default}~\cite{the_european_union_regulation_2016}. However, research has shown, for example, that simply removing user identifiers is insufficient for anonymizing data adequately~\cite{narayanan_robust_2008} and especially concerning mobility data, 4 randomly known data-points are enough to identify 95~\% of mobility traces uniquely~\cite{de_montjoye_unique_2013}.

DP and other PPTs are good candidates to match the technical requirements of data protection and they are already used in practice~\cite{abowd_us_2018, erlingsson_rappor_2014, ding_collecting_2017}. Especially DP provides a unique opportunity of a mathematically proven guarantee in form of an upper bound to the privacy risk, of sensitive data being exposed. When using DP this guarantee is an essential piece of information for the sharing decision process. Therefore, easy to understand quantitative privacy risk notifications are needed.

The design and impact of privacy notices in general is a widely researched field of work (\eg~\cite{schaub_designing_2017,bohme_security_2011,felt_android_2012}). While previous research has produced many recommendation on how to communicate qualitative privacy information and even qualitative privacy risk~\cite{rajivan_influence_2016}, an effective communication of the quantitative privacy risk information of DP is still missing.

Previous attempts to explain DP in particular have led to mixed feedback.
The US Census Bureau, for example, experienced that people did not understand the functioning of DP\footnote{\null\url{https://nypost.com/2021/08/26/elsie-eiler-surprised-when-one-person-nebraska-town-doubles-in-census}\null, accessed 2021-10-06.} or even the necessity for using DP\footnote{\url{https://www.nytimes.com/2018/12/05/upshot/to-reduce-privacy-risks-the-census-plans-to-report-less-accurate-data.html}, accessed 2021-10-06.}.
This problem has inspired first academic investigations into the best strategy for communicating DP to laypeople, which we summarize in the following.

Bullek et al.~\cite{bullek_towards_2017} conducted a study to explain
DP using the metaphor of randomized response technique~\cite{warner_randomized_1965}.
Instead of modifying the data after collection, the randomized response technique instructs participants to randomize their answers directly according to a \textit{Wheel of Fortune}-like visualization:
the participants either answer truthfully or answer ``Yes'' or ``No'' as indicated on the randomized visualization regardless of their true answer. 
The distribution of \enquote{answer truthfully}, \enquote{answer Yes}, and \enquote{answer No} on the wheel can be adjusted to tune the privacy protection in accordance with the DP parameter $\varepsilon$.
Based on this procedure, participants were able to correctly rank different DP options according to their privacy guarantee.
These results suggest that providing individuals with an intuition on privacy risks is possible and beneficial for the individual's decision-making.
However, this positive result has two caveats. 
Firstly, Bullek et al. only measured the ability to compare multiple DP settings.
We focus our research on the more common situation, where individuals have to decide whether a single privacy guarantee is strong enough to share personal data. 
Secondly, as noted by Bullek et al., participants perceived the procedure as lying when prompted to simply answer \enquote{Yes} or \enquote{No.} This influenced individuals in the study towards choosing a less private option against their privacy preference to avoid lying. By focusing on the privacy guarantees instead of the mechanism that achieves the guarantee, this effect should be avoided.

Xiong et al.~\cite{xiong_towards_2020} compared different real-world notifications about DP.
They examined, for example, whether certain descriptions of DP are easier to understand or whether a positive framing of DP, \ie its privacy guarantee, leads to different results compared to a negative framing of DP, \ie the privacy risk.
The researchers recognized that several unintentional differences in descriptions (\eg the mention of a known institution as data collector) had a stronger influence on individual's decision-making than the influence of the differences in description. The addition of quantitative risk descriptions with their objective nature might have the potential of a less volatile effect.

Cummings et al.~\cite{cummings_i_2021} investigate the effect of the theme (\eg techniques, trust or risk) of various real-world notifications about DP on individuals' willingness to share data. 
They did not find any significant effects on the decision-making, postulating that none of the descriptions provided a meaningful enough mental model of DP. 
Nonetheless, they identified the descriptions that centered around privacy risk as the most promising, because this description conveyed the most accurate privacy expectations to individuals. 
As the privacy guarantees of DP describe a bound on the privacy risk, this result also sound promising for our proposed communication.

In summary, the existing research on how to explain DP to laypeople supports the claim that communicating DP is challenging. Even developers and technically skilled people face the challenge of understanding  
DP in all its details~\cite{hsu_differential_2014,dwork_differential_2019}. 

Therefore, we take a different approach and consider what transparency means in the context of DP. We argue, consistent with the results by Cummings et al.~\cite{cummings_i_2021}, transparency with respect to DP should focus on the privacy risk and specifically on the offered guarantees as described by the DP definition.
In particular, our approach is to translate the guarantee into a quantitative risk communication format derived from the actual value chosen for the privacy budget $\varepsilon$.
As the choice of $\varepsilon$ heavily influences the resulting protection, only with this transparency DP can be a useful tool in supporting individuals in privacy decision situation.

However, communicating quantitative privacy risk has been explored little in privacy research. We rely, therefore, on risk communication in the medical domain which is a well-developed field of research. In the following, we provide an overview of existing medical risk communication formats.

\subsection{Medical Risk Communication}
\label{sec:communicatingrisk}

The goal of medical risk communication is to enable an individual (here a patient), to make an informed decision about a particular treatment.
There are a number of meta-studies that offer recommendations on how to communicate medical risks effectively which we used for reviewing existing medical risk communication formats~\cite{gigerenzer_helping_2007, visschers_probability_2009, trevena_presenting_2013, bansback_communicating_2017, bonner_current_2021}\footnote{We used Google Scholar with search terms \enquote{health decisions risk best practices review} and \enquote{health probability presentation best practices,} restricted to review papers and the last 15 years and manually screened the first 200 papers each by relevance.}.
Based on our analysis, we decided to focus on communicating aleatory 
uncertainty, \ie uncertainty about future outcomes, instead of epistemic uncertainty, \ie uncertainty due to the lack of information.
Furthermore, we do not consider 2-variate risk formats (\eg natural frequencies)
because they do not support our goal of explaining DP.
From the meta-studies, we derived six basic risk communication formats which are often used in medical risk communications: \textit{percentages}, \textit{simple frequencies}, \textit{fractions}, \textit{numbers needed to treat}, \textit{1-in-x} and \textit{odds}.
\footnote{We provide a detailed overview of the risk formats identified in
\autoref{sec:riskFormatsDetails}}. 
Some of these formats are generally discouraged, since they can be misunderstood easily (\emph{1-in-x}. \emph{numbers needed to treat}), they are inflexible in use or inaccurate (\emph{fractions}), or not well researched (\emph{odds}).

The two remaining base formats \textit{percentages} and \textit{simple frequencies}, are both recommended in different settings.
In addition, we identified several format variations that can be added to these basic formats: \textit{comparison to known risks}, \textit{comparison to peers}, \textit{comparison to status quo} and \textit{outcome framing}.
In the following we briefly introduce the two suitable base formats and the variations.

\emph{Percentages} are a widely used and well-known risk communication format (\eg \enquote{23.8~\% of patients experience dizziness})~\cite{trevena_presenting_2013, ancker_covid-19_2020, bonner_current_2021}. Travena et al.~\cite{trevena_presenting_2013} recommend this format especially when comparing multiple risk-related options. McDowell et al.~\cite{mcdowell_simple_2016} only caution about presenting risks lower than 1~\% in this format. For all other values, rounding to the nearest whole number is generally recommended.

%
\emph{Simple frequencies}~\cite{gigerenzer_helping_2007, mcdowell_simple_2016, trevena_presenting_2013, bonner_current_2021} presents the number (numerator) of cases affected by the event compared to a meaningful baseline (denominator) (\eg \enquote{13 out of 350 patients experience dizziness}). This format is commonly accepted and understood by laypeople.
However, multiple options exist for the numerator and the denominator to represent the same risk and especially the choice of a meaningful denominator might affect the efficiency. There are multiple complementary recommendations on how to make an 
appropriate choice:
Several of the meta-studies strongly recommend using consistent denominators to minimize the mental load on the individuals when comparing different risks in this format. 
Multiples or even powers of 10 as denominator have been shown to be beneficial~\cite{brick_risk_2020}. 
Small denominators are recommended because they are better understood and easier to remember.
Concerning the numerator, there is strong evidence that $1$ as a numerator skews the risk perception of individuals~\cite{sirota_1--x_2018} and, therefore, higher values are recommended.

In addition the meta-studies discuss the following variations, which can be added to the basic formats.
%
Firstly, a valuable information is the comparison to the risk of a similar event~\cite{gigerenzer_helping_2007,bonner_current_2021} (\eg \enquote{The risk of side effects due to radiography is 15~\%, only slightly higher than the 13~\% risk due to other common radiation sources}).
%
Secondly, the risk of an event can also be compared to the risk of the same event for peers~\cite{bonner_current_2021} (\eg \enquote{The risk of suffering a heart attack is 15~\%, which is higher than the 13~\% of the average US American}).
%
Finally, the risk of an outcome with an action can also be compared against the risk of the outcome without the
action, \ie the status quo~\cite{gigerenzer_helping_2007,trevena_presenting_2013}. This is often used to explain the reduction of risk due to a treatment (\eg \enquote{Treatment A leads to a survival chance of 60~\% compared to a survival chance of 40~\% without the treatment}). 

Another common variation is the \textit{outcome framing}. To avoid bias due to presenting the positive or the negative effect (\enquote{5~\% of participants have
side effects} or \enquote{95~\% do not have side effects.})
several meta-studies (\eg~\cite{spiegelhalter_risk_2017,spiegelhalter_visualizing_2011})
recommend presenting the probabilities of both events together. \enquote{5~\% have
side effects, 95~\% do not.}


Overall, results of medical risk communication research show that the success of an approach depends on a multitude of factors including situational factors (\eg decision between two options versus the decision for or against a single option, the range of the presented risk) and personal factors (\eg experience with statistics). When transferring risk communication formats to the domain of privacy risks, we, therefore, need to re-evaluate the formats in the new context.


\section{Communicating Privacy Guarantees}
\label{sec:design}
In \autoref{sec:communicatingrisk} we identified risk communication formats from the medical domain. These formats are well researched in the medical domain and, therefore, good candidates to present privacy risks in quantitative privacy risk notifications, which communicating the privacy guarantees of DP.

As a first step into such quantitative privacy risk notifications this study investigates the following two research questions:

\setlist[itemize]{leftmargin=7.5mm}
\begin{itemize}
    \item [RQ1] How do established quantitative risk communication formats from the medical domain perform in terms of subjective and objective understanding of privacy risk, compared to existing qualitative privacy risk notifications?
    \item [RQ2] How is the effectiveness of privacy risk communication influenced by personal attributes of individuals, such as gender, statistical numeracy and privacy aptitude?
\end{itemize}

In \autoref{sec:RiskCommunicationFormats}, we discuss suitable risk communication formats from the medical domain, in \autoref{sec:representingepsilon} we explain how the guarantees provided by DP can be presented in risk communication formats and, in \autoref{sec:notifications}, we describe how the formats were incorporated into privacy risk notifications.

\subsection{Selecting Risk Communication Formats}
\label{sec:RiskCommunicationFormats}

The decision situation in the medical domain is similar to the privacy domain:
patients have to decide whether to undergo a treatment which changes the probability of an adverse outcome. The change in probability of the adverse outcome is specified in a quantitative risk format but the result of the treatment remains uncertain. 
Therefore, risk communication formats used in the medical domain can be used to represent privacy risks.

As discussed, two formats are recommended as best practice formats in different medical contexts, the \emph{percentage format} and \emph{simple frequency format}.
In the following, we discuss the adaptation of the formats into the privacy domain and also consider the recommended variations.


The \emph{percentage formats} are well-known and, therefore suitable for presenting privacy guarantees. An example of presenting a privacy guarantee in this format is: \enquote{\textit{When using DP, your risk of being identified in the dataset is at most 52~\%}.}
As recommended, we avoid decimal point by rounding to the nearest whole number. Since DP calculates worst-case upper bounds on the privacy risk of all individuals, the resulting privacy guarantees are never below 1~\%, circumventing the known issues of the percentage format in this range.

The format \emph{simple frequencies} is also suitable to represent privacy risk.
An example that applies this format in the privacy domain is
\enquote{\textit{When using DP, at most, 26 out of 50 participants are identified in the dataset.}}
Weighing the recommendations, we use multiples of 10 over powers of 10 in our representations to avoid unnecessary high denominators or inaccuracy due to rounding. Furthermore, we prioritize avoiding $1$ as numerator over small denominators due to the strong evidence against its understandability.

In contrast to the medical domain, time frames are not relevant in DP privacy guarantees as all information is considered to be known from the date of publication. The individual is either identifiable immediately or remains anonymous.

Concerning the variations identified in \autoref{sec:communicatingrisk} we consider \emph{outcome framing} and \emph{comparison to status quo}, since they relate well with DP.
In both base formats, \emph{percentages} and \emph{simple frequencies}, we usually present the probability of the negative outcome, \ie the risk of being identified when sharing the personal data.
The variation \emph{comparison to status quo} adds the ground-probability of being identified before sharing data. This variation is naturally related to DP, since DP compares by definition the probabilities in the two described situations.
In a second variation, \emph{outcome framing}, we additionally present the counter-probability, \ie the probability that the re-identification does not occur.
This variation is especially well suited, in the context of DP, since the mathematical definition of DP is formulated in terms of a privacy guarantee rather than a risk.

The other two comparison variations cannot be translated easily to the privacy domain.
The concept of risk of an average person, as used in the variation \emph{comparison to peers}, is difficult to translate into the privacy domain, as according to Bhatia et al.\cite{bhatia_empirical_2018} privacy risk is highly subjective and can only be measured by unacceptably sever surveillance.
Furthermore, for the variation \emph{comparison to similar event}, the outcome of a possible privacy leak cannot be easily compared to any other well-known risky event because data for such a comparison are not available.

In summary, we selected two base formats: the \emph{percentages format} 
and the \emph{simple frequencies format}. We can use each base-format as-is or with either of the two variations \emph{comparison to status quo} or \emph{outcome framing}. 
Thus, we arrive at six (2x3) possible quantitative privacy risk communication formats.

Next we discuss possible approaches for translating a privacy budget $\varepsilon$ into one of the identified risk communication formats.

\subsection{Representing DP guarantees}
\label{sec:representingepsilon}

While a lot of work has addressed the mathematical mechanisms of DP, there is less research on how to interpret $\varepsilon$ as a common risk ~\cite{naldi_differential_2015,hsu_differential_2014,tsou_rod_2021}.

One possible approach for such a translation is provided by the model of Lee and Clifton~\cite{lee_how_2011}. They rephrase $\varepsilon$ as a probability of identifying any particular individual as having contributed to the result of the analysis depending on two additional factors: the number of individuals in the data set and the maximal impact one individual could theoretically have on the outcome of the query.
Mehner et al.~\cite{mehner_towards_2021} simplified this model by considering the worst-case bound for the additional parameters.
Thus, their notion of \emph{global privacy risk} $P$ provides a translation of $\varepsilon$ into the desired percentage or frequencies format:
\begin{align*}
    P = \frac{1}{1+e^{-\varepsilon}} \text{.}
\end{align*}

The advantage of the model by Mehner et al., in contrast to Lee and Clifton~\cite{lee_how_2011} and other models, is that the obtained values are sound privacy guarantees, independent of any parameters, which are unknown at the time of data collection. Therefore, this model is a good fit for the goal of this study.

\subsection{Privacy risk notifications}
\label{sec:notifications}
\begin{figure}[tb]
    \centering
    \includegraphics[width=\columnwidth]{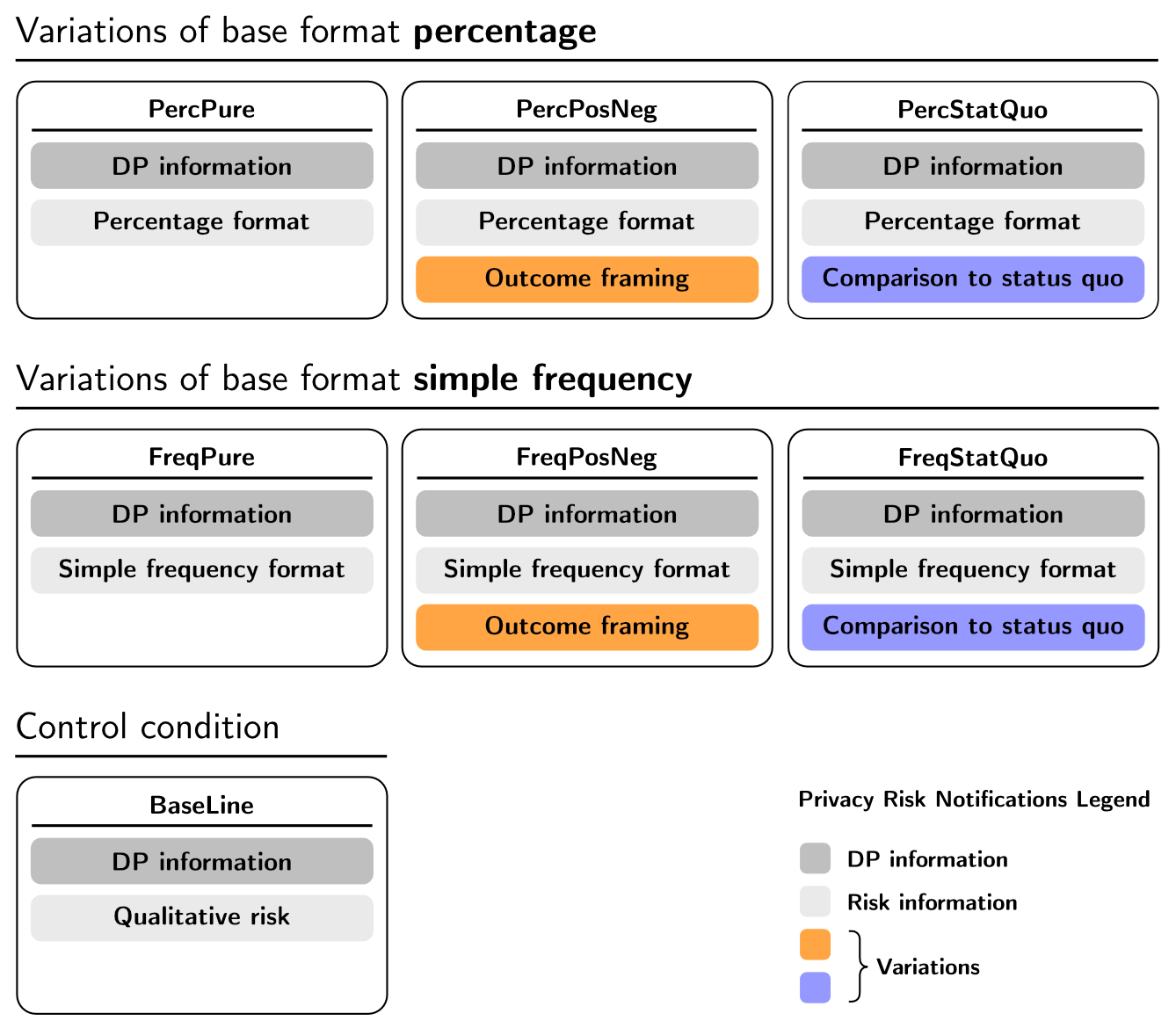}
    \Description[Overview of the privacy risk notifications]{The two base risk formats percentage and frequency are each present in 3 variations: Pure, StatQuo and PosNeg. The latter two contain the pure variation but each add more information at the end. The seventh notification is the notification BaseLine}
    \vspace{-2.5em}
    \caption{Overview and composition of the seven privacy risk notifications.}
    \vspace{-2em}
    \label{fig:con}
\end{figure}

Using the DP model, we can calculate the DP guarantee as a worst-case privacy risk and represent the resulting values using the six identified risk communication formats discussed in \autoref{sec:RiskCommunicationFormats}. These six formats are used to specify six novel quantitative privacy risk notifications, which are then evaluated. In addition, as notification \emph{BaseLine} we use the best performing qualitative privacy risk notification from Xiong et al.(cf. notification \enquote{DP Imp.} in \cite{xiong_towards_2020}). Thus, overall, we consider the seven privacy risk notifications in \autoref{fig:con}.

Each privacy risk notification contains the same general DP information (extracted from the BaseLine notification), information about the privacy risk (in qualitative form or in one of the two quantitative base formats) and potentially additional information dependent on the variation.
The wording used to incorporate the base formats was based on similar medical risk notifications compiled by Bansback et al.~\cite{bansback_communicating_2017} and developed in multiple iterations with experts and laypeople on DP. We aimed for mathematical correctness of the notification and simultaneously an understandable description for general users.

To match the rather strong privacy guarantees suggested in the wording of the \emph{BaseLine} notification, we used the value $\varepsilon=0.1$ as privacy budget. With this value and the model by Mehner et al.~\cite{mehner_towards_2021}, we calculated the privacy guarantees $P=0.525$, which results in the notifications \enquote{With differential privacy, at most 26 out of 50 | 52~\% of statistics will reveal whether you visited a location.} 

These values also determine the counter-probability of not being identified, as used in the variation \emph{outcome framing}.
In the variation \emph{comparison to status quo} we use the worst-case value $P_{guess}=0.5$ as outlined by Mehner et al.
This value describes the probability of correctly guessing the presence of an individual in the data set without information from the data set, which corresponds to the status quo, before sharing one's data.
We present this probability as a percentage \enquote{50~\%} and a frequency \enquote{25 out of 50}.

It is worth noting that all calculated risks present realistic values according to the chosen privacy budget.


\section{Experimental Study}
\label{sec:studyDesign}

Based on the privacy risk notifications, we designed and conducted an experimental study on Amazon Mechanical Turk (MTurk) in order to evaluate the suitability of our approach and understand the differences in effectiveness between the selected formats. 
We decided for a between-subject study to avoid potential learning effects.

\subsection{Study Design}
\begin{figure*}[ht]
    \centering
    \vspace{-1em}
    \includegraphics[width=\textwidth]{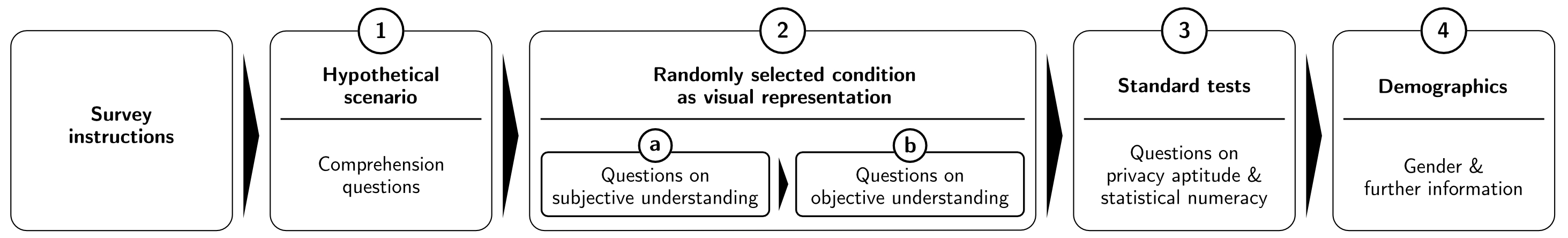}
    \vspace{-2.4em}
    \caption{Overview of the study design including the four phases.}
    \vspace{-1em}
    \Description[Visualizations of the four phases of the study]{From left to right we have the phases: survey instructions, 1: hypothetical scenario containing comprehension questions, 2: Randomly selected condition as visual representations containing questions on (a) subjective and (b) objective understanding, 3: Standard tests on privacy aptitude and numeracy and 4: Demographics including gender and further information.}
    \label{fig:studyproc}
\end{figure*}

Our study setup was divided into four main phases (cf.~\autoref{fig:studyproc}): Participants
(1) were introduced to a scenario with a focus on mobility data;
(2) were shown one of seven conditions at random, \ie a privacy risk notification, and seeing the notification 
(2a) were asked to self-rate their understanding of
the notification;
(2b) were asked to answer objective true/false questions
about the privacy risks of DP;
(3) completed standard tests about their privacy aptitude
and their statistical numeracy, and, finally, 
(4) were asked to indicate their perceived gender.
We describe the design consideration for each phase in the remainder of this section. All study resources can be found in \autoref{sec:studyResources}.

We defined a fictional ridesharing scenario to capture the complexity of privacy risks.
Ridesharing services show increasing popularity and, therefore, have a high likelihood of being relatable by the participants. The scenario describes that, after an app update, user data from the ridesharing app will now be shared with a local authority to improve the urban infrastructure. For simplicity, the scenario does not include a privacy decision. Our research goal is concerned with establishing the requirements for an informed decision, rather than facilitating the decision itself. The scenario text was refined in a thinking-out-loud session with English native speakers.

In the context of this scenario, the participants were exposed to one of the privacy risk notifications, presented as a typical pop-up design on a stylized smartphone (cf. \autoref{fig:condition}).
We then evaluated the participants' subjective and objective understanding of the privacy guarantees:
for the subjective understanding we asked participants to 
rate (S1) their perceived general understanding of the notification, (S2) their perceived ability to assess individual privacy risks based on the notification and (S3) whether they felt the descriptions provided all necessary information each on a 7-point Likert scale.
Concerning the objective understanding, we asked participants to judge true/false statements about DP. We compared the privacy risk of DP with the risk (O1) when sharing the unprotected data and (O2) when sharing no data at all. In addition, we asked about two basic principles of DP: (O3) the assumption, that a basic privacy risk exists without sharing more data and (O4) that the distribution of the privacy risk cannot be manipulated by the data consumer, to reveal one specific target location.
Where possible, we randomized the correct answer (true/false) of these questions by varying the qualifier (less/more, higher/lower).

%
In RQ2 we consider a number of context factors, which have shown an influence on risk communication in other contexts.
An important influence for the effectiveness of risk communication is the ability of the individual to handle statistics and probability information in general. This attribute is commonly known as \emph{(statistic) numeracy}. In our study we measure the numeracy of our participants with the \textit{Berlin Numeracy Test}~\cite{cokely_measuring_2012} in its adaptive format.
%
Research also shows an influence of gender, especially on the perception of positive or negative 
framing of risk communication~\cite{huang_sex_2010}. 
We, therefore, ask participants to self-report their gender following recommendations by Spiel et al.~\cite{spiel_how_2019}.
%
Finally, privacy aptitude is an obvious potential influence. We adapted the \textit{Internet Users' Information Privacy Concerns} test (IUIPC)~\cite{malhotra_internet_2004} to evaluate the privacy aptitude of participants. This test divides the privacy aptitude into the three dimensions \enquote{control,}\footnote{How important is control of personal data and autonomy?} \enquote{awareness}\footnote{How important is the knowledge concerning how personal data is collected, processed and used?} and \enquote{collection}\footnote{How concerned is the participant with data collection?} 
each with three to four questions. Since the participants did not have control over the data in our scenario, we measured only for \enquote{awareness} and 
\enquote{collection,} which are both directly relevant to our scenario.

We included comprehension check questions after the scenario and attention check questions as part of the objective understanding, and the privacy aptitude test to exclude inattentive participants.
At the beginning of the study, we informed the participants that MTurk IDs and IP addresses would be stored for quality control purposes, but would be deleted once quality was verified. 
We also explained the collection of gender information to the participants and its purpose in the study.

\subsection{Participant recruitment and study execution}

Although MTurk has known disadvantages concerning nuanced representation of users~\cite{aguinis_mturk_2021}, its advantage of reaching a wide audience in this early investigation into the suitability of the risk communication formats outweighs these concerns.
%

We restricted participants to MTurk workers with at least 95 \% approval rating and 
at least 1,000 approved assignments.
As customary, we restricted participants to US-residents, to increase the likelihood of an adequate English proficiency.
We determined a working time of about 12 minutes, 
with careful reading of all material, in a small pre-test run with a few selected participants (recruited from our research group as well as project partners). Based on a conservative working time of 15 minutes, each participant was paid \$3.00 (\$12.00/hour). This is 
significantly higher than the local minimum wage (\textasciitilde\$10.01/hour at the time of recruitment) and about equal to the average hourly wage on MTurk (\textasciitilde\$12.05/hour)\footnote{\null\url{http://faircrowd.work/platform/amazon-mechanical-turk},  
accessed 10.12.2021}.
However, we allowed up to 40 minutes per participant to avoid automatic timeouts of participants who were slower for any reason.
The MTurk assignments were distributed throughout the day between reasonable US working hours.


\section{Statistical Analysis}
\label{sec:results}

In this section, we present the procedure and the results of our statistical analyses\footnote{We provide all anonymized data and our R scripts under open access on OSF upon acceptance of the article.}. We start with 
the general preprocessing and insights into the collected data set, then report on the analysis regarding RQ1 and RQ2.
Throughout the analysis, we use a significance-level of $\sigma=0.05$ and adjust the results of the post-hoc t-tests with Bonferroni-corrections\footnote{t-tests compute many correlations on the same data set. By the nature of these correlations, the likelihood of discovering false correlations increases with the number of computed correlation values. Bonferroni is one of the common practices to correct the resulting values for this effect.}. 

\subsection{Data set and Pre-processing}
Based on an \emph{a priori} power analysis (f=.25, power=.8), we aimed for a sample size of 338 submissions and collected a total of 444 submissions. We excluded submissions for failed attention or comprehension check questions (95) and repeated participation (6). 
We are confident, that we eliminated most unreasonable outlier with these test and, therefore, decided against further elimination of outliers based on values.
The cleaned data set contains 343 submissions, which are approximately equally distributed between the seven groups (cf.~\autoref{fig:SubjUnderstanding_avg}). Average working time was around 8:40 minutes.

The three answers about subjective understanding were combined into its arithmetic mean. The resulting values lie between 1 and 7, with high values denoting that the participant felt confident in their understanding.
For the objective understanding we marked each correct answer with 1 point. The summed score for all four answers constitutes the objective understanding value between 0-4.

Out of the 343 participants, 117 as female (F) and 223 reported themselves to be male (M). With 65.0 \% male participants the ratio is slightly higher than similar studies~\cite{mccredie_who_2019}. We combined the three participants reporting being non-binary or self-described into a third category (X).

According to the BNT, we categorized submissions into four numeracy groups\footnote{The Berlin Numeracy Test (BNT)~\cite{cokely_measuring_2012} provides a flow-diagram to classify participants into one of four levels}.
The results were highly skewed towards low numeracy (\emph{lowestN}: 175, \emph{lowN}: 47, \emph{highN}: 39, \emph{highestN}: 82). The test is designed to result in four equal groups with university-affiliated participants. 
However, the lower rate of MTurk workers with a university degree (around 49~\% \cite{moss_is_2020,warberg_can_2019}) plausibly explains the difference to the BNT evaluation study~\cite{cokely_measuring_2012}.

\begin{figure}
    \centering
    \includegraphics[width=\columnwidth]{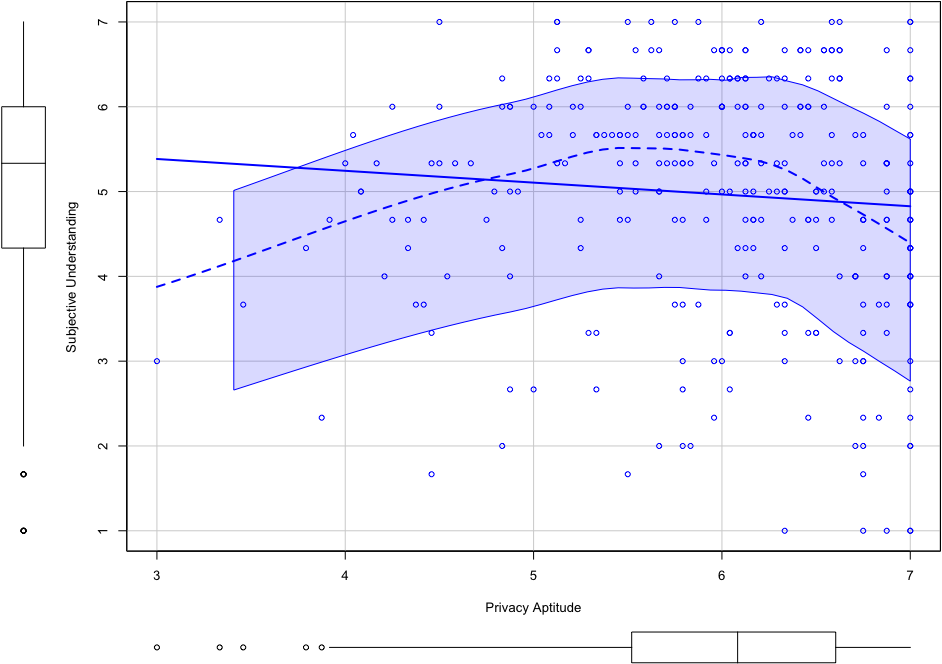}
    \Description[Scatterplot with privacy aptitude (x-axis) against subjective understanding (y-axis)]{The plot contains a modelling of the relation. It starts at aptitude 3 with a value slightly below 4, and increases until aptitude 5.3 with a value of 5.5, until aptitude 6.2 it stays around value 5.5 and then declines to a value of 4.5 for aptitude 7}
    \vspace{-2em}
    \caption{Privacy aptitude against Subjective Understanding}
    \vspace{-1em}
    \label{fig:PrivacyAptitude_vs_SubjectiveUnderstanding}
\end{figure}

We used the results from the privacy aptitude questionnaire, to compute separate arithmetic means for \enquote{awareness} (three questions) and \enquote{collection} (four questions). 
We then calculated the arithmetic mean of these two values to get an aggregated value for the privacy aptitude\footnote{For example, if a participant answered (5,3,7) on awareness  and (3,6,2,5) on collection, we calculated $\mathrm{awareness}=\nicefrac{(5+3+7)}{3}=5$, $\mathrm{collection}=\nicefrac{(3+6+2+5)}{4}=4$, and finally $\mathrm{privacy aptitude}=\nicefrac{(5+4)}{2}=4.5$.} on a scale between 1 and 7.
Additionally, we sorted submissions into privacy aptitude groups (cf.~\autoref{sec:covariates_model}). By visual inspection of the relationship privacy aptitude to subjective understanding (\autoref{fig:PrivacyAptitude_vs_SubjectiveUnderstanding}), we identified rising values in subjective understanding up to a privacy aptitude value of $~5.3$, a plateau up to $~6.2$, and falling subjective understanding afterwards. According to those three different areas we categorized the privacy aptitude into three groups \emph{lowP}, \emph{mediumP}, and \emph{highP}.

In the final data set each entry contains the following values: condition group \{1-7\}, subjective understanding [1-7], objective understanding [0-4], Numeracy \{\emph{lowestN}, \emph{lowN}, \emph{highN}, \emph{highestN}\}, privacy aptitude group \{\emph{lowP},\emph{mediumP},\emph{highP}\} as well as the accurate privacy aptitude value [1-7] and gender \{F,M,X\}.

\subsection{The Effect of Risk Communication Formats on Subjective and Objective Understanding}
\label{subsec:effectrisformat}

\subsubsection{Statistical Model and Requirements}
In RQ1 we investigate the influence of the shown notification (\eg \textit{FreqPure}) on subjective and objective understanding. 
To answer RQ1 we investigate ANOVA-models, with the condition group as independent variable and the objective or respectively the subjective understanding as dependent variable. In both models $H_0$ is \enquote{The groups performed equally.}

In the case of subjective understanding we use a Welch-ANOVA, to account for dissimilarity of the variances (Levene test $p = 0.0466$). The data on the objective understanding allows a standard ANOVA. All other requirements have been checked without issue.
To investigate potential differences, we use a post-hoc t-test.

\subsubsection{Results}
The means of the subjective understanding by condition (cf.~\autoref{fig:SubjUnderstanding_avg}) show that the \textit{BaseLine} notification ($5.6$) performed better than all quantitative privacy risk notifications ($4.58$ - $5.11$).
The Welche-ANOVA on the subjective understanding confirmed a significant main effect ($p = 0.009786$), indicating that the notifications had an influence on the subjective understanding.
The post-hoc t-test, (cf.~\autoref{fig:SubjUnderstanding_avg}) reveals significance only between the notifications \emph{BaseLine} ($5.6$) and \emph{FreqStatQuo} ($4.58$).
\begin{table}[tb]
    \centering
    \scriptsize
        \begin{tabularx}{\columnwidth}{@{}llllc@{\hspace{2px}}lllllll@{}}
            \toprule
            & & \multicolumn{2}{c}{Aritm. Mean} & &\multicolumn{6}{c}{Post-hoc t-test (Subj. Underst.)} \\ \cline{3-4} \cline{6-11}
            &\begin{tabular}[c]{@{}l@{}}Size\end{tabular}& \begin{tabular}[c]{@{}l@{}}Subj. \\Underst.\end{tabular} & \begin{tabular}[c]{@{}l@{}}Obj.\\ Underst.\end{tabular} & & (1)    & (2)    & (3)    & (4)    & (5)    & (6)      \\\midrule
            FreqPure    \hfill(1)&42& 4.88  & 1.69&& -&-&-&-&-&-\\
            FreqPosNeg  \hfill(2)&59& 5.05  & 1.63&& 1& -& -& -& -& -    \\
            FreqStatQuo \hfill(3)&46& 4.58  & 1.98&& 1& 1& -& -& -& -    \\
            PercPure    \hfill(4)&46& 5.07  & 1.61&& 1& 1& 1& -& -& -    \\
            PercPosNeg  \hfill(5)&57& 5.11  & 1.88&& 1& 1& 1& 1& -& -    \\
            PercStatQuo \hfill(6)&45& 4.99  & 1.58&& 1& 1& 1& 1& 1& -    \\
            BaseLine    \hfill(7)&48& 5.60  & 1.79&& .28& .80& .0074$^{**}$& 1& 1&  .70 \\\bottomrule
        \end{tabularx}
    \caption*{\small Mean subjective and objective understanding of participants in each condition and pairwise t-test (Subj. Understanding)\\$^{**}p\le 0.01$, $^*p\le 0.05$, $^+p\le 0.1$}
    \vspace{-2em}
    \caption{Effects of Formats}
    \label{fig:SubjUnderstanding_avg}
    \vspace{-4em}
\end{table}

Concerning objective understanding (cf.~\autoref{fig:SubjUnderstanding_avg}), we identify the best performing notifications to be \emph{FreqStatQuo} ($1.98$) and \emph{PercPosNeg} ($1.88$), both performing better than the \emph{BaseLine} notification ($1.79$). However, the difference is not shown to be significant by the standard ANOVA test~($p=0.3053$). Consequently, we cannot claim a significantly better performance of the two mentioned quantitative risk communication formats.

\subsection{The Effects of Individuals' Characteristics}
\label{sec:results_covariates}

\subsubsection{Statistical Model and Requirements}
\label{sec:covariates_model}
In \autoref{sec:studyDesign} we hypothesized that numeracy, privacy aptitude or gender might have an influence on the effectiveness of the privacy risk notifications, \ie on the subjective and objective understanding. 

To evaluate our hypothesis, we constructed a linear regression model each for subjective and objective understanding. 
We include seven independent predicates for each model: (1) condition group, (2) numeracy, (3) privacy aptitude group, (4) gender, and three interaction terms (5) group:numeracy, (6) group:privacy aptitude group and (7) group:gender. The dependent variable is the objective or respectively subjective understanding.
Since the relationship between privacy aptitude and objective/subjective understanding is not linear, we use the privacy aptitude groups in the linear regression model instead.

As base categories for the linear model, we used the \textit{BaseLine} condition, the highest numeracy level \emph{highestN}, the medium privacy aptitude \emph{mediumP}, and the gender \emph{male} \footnote{These base categories define the group to which differences are calculated. We aim to discover differences caused by the risk communication formats. Thus, we choose the \textit{BaseLine} condition for comparison. Since we expected higher differences in lower levels of numeracy, we selected the highest level of numeracy to discover differences to the lower levels. For privacy aptitude we did not have an intuition and, therefore, chose the medium level to discover effects of the levels low and high simultaneously. The choice for gender base category is irrelevant with only two considered values. We kept the value \emph{male} for gender as defaulted to by the linear regression model.}

\subsubsection{Results}
In the linear regression model for the subjective understanding, we observe significant effects for \emph{PercStatQuo} ($p=0.0255$) and for the interaction between the condition \emph{PercStatQuo} and the numeracy \textit{lowN} ($p=0.0248$). The table of all regression coefficients is shown in \autoref{sec:statistic_resources_subjective}.

\begin{figure}
    \centering
    \begin{subfigure}[b]{\columnwidth}
        \includegraphics[width=\columnwidth]{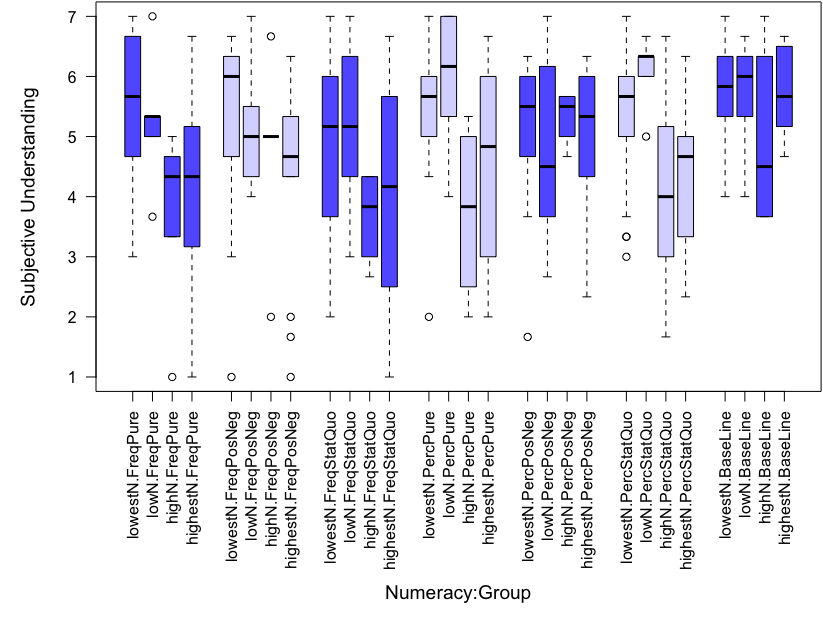}
        \vspace{-3em}
        \Description[Box plot showing the subjective understanding for the numeracy levels for each condition]{Each condition is divided into four vertical bars showing the result of participants with lowest, low, high and highest numeracy level. For the conditions FreqStatQuo, PercPure, PercStatQuo and BaseLine the pattern from the text can be seen.}
        \caption{Effect of Numeracy on Subjective Understanding}
        \label{fig:SubjUnderstanding_group_and_numeracy}
    \end{subfigure}
    \begin{subfigure}[b]{\columnwidth}
        \includegraphics[width=\columnwidth]{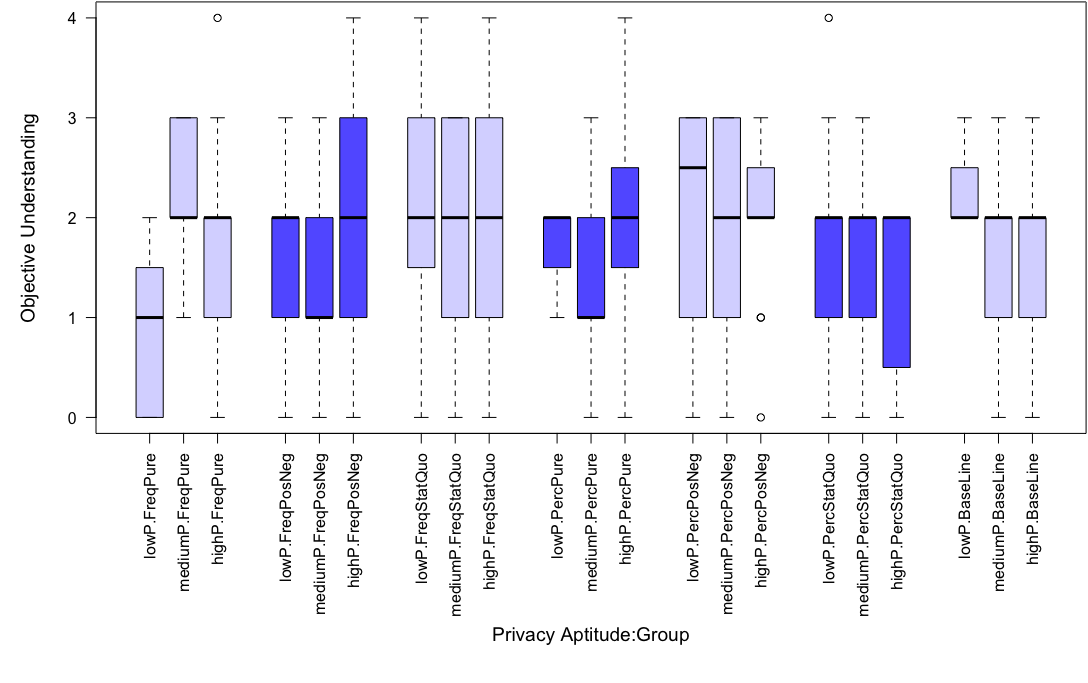}
        \vspace{-2.5em}
        \caption{Effect of Privacy Aptitude on Objective Understanding}
        \Description[Box plot showing the subjective understanding for the privacy aptitude levels for each condition]{Each condition is divided into three vertical bars showing the result of participants with low, medium, high privacy aptitude. All bars are centered around the value 2 or slightly below, only the bar for FreqPure with low privacy aptitude is lower centered around 1.}
        \label{fig:ObjUnderstanding_group_and_privacy}
    \end{subfigure}
    \vspace{-2em}
    \caption{Effect of attributes}
    \vspace{-2em}
\end{figure}

\autoref{fig:SubjUnderstanding_group_and_numeracy} shows the subjective understanding depending on condition group and numeracy. 
The diagram reveals that the \textit{PercStatQuo} notification (the right most light blue group) has one of the lowest results in the sub-groups \textit{lowestN}, \textit{highN} and \textit{highestN} (labeled~\enquote{\textit{lowestN.PercStatQuo}}, \enquote{\textit{highN.PercStatQuo}}, \enquote{\textit{highestN.PercStatQuo}}), but the highest medium performance in the subgroup \emph{lowN} (labeled~\enquote{\emph{lowN.PercStatQuo}}). 

Furthermore, the four numeracy groups of the \textit{PercStatQuo} notification show a distinct pattern; the subjective understanding is moderate in the lowest numeracy level \emph{lowestN}, then it continues with a higher value in the low numeracy level \emph{lowN}, followed by a low value for the high numeracy level \emph{highN}, and finally again with a moderate value in the highest numeracy level \emph{highestN}. 
This pattern can also be seen for the \textit{BaseLine}, \textit{PercStatQuo} and \textit{PercPure} notifications, though the effect with these notifications was not shown to be significant.

The equivalent linear regression model for objective understanding (cf.~\autoref{sec:statistic_resources_objective}) returns a significant effect ($p=0.033435$) only for participants with low privacy aptitude \emph{lowP} and the \emph{FreqPure} notification. 
In \autoref{fig:ObjUnderstanding_group_and_privacy} a box plot shows that these participants performed worst in objective understanding. However, this is an isolated finding which cannot be observed in any other group.


\section{Discussion}
\label{sec:discussion}
\label{sec:Discussion_Results}

The overarching goal of our research is to enable laypeople to share their data for the benefit of the public in a privacy-preserving manner by employing DP. The guarantees offered by DP are mathematically sound, but to support laypeople in decision situations, these guarantees need to be communicated in a transparent and understandable way.
To support this goal, we investigated the understandability of quantitative privacy risk notification. In the following section, we reflect on the results of our study. 

We discuss our results from four angles. Firstly, we highlight the general suitability of using quantitative privacy risk notifications. Secondly, we reflect on how the risk communication formats influence subjective and objective understanding, thirdly, we discuss the influence of numeracy on the effectiveness of the formats and finally we summarize our thoughts on the influence of the other attributes.

\subsection{Suitability of Risk Communication Format}
\label{sec:disc_riskformat}

We enhanced existing privacy risk notifications from existing usable security and privacy research (cf.~\cite{cummings_i_2021,xiong_towards_2020}) with quantitative risk communication formats from medical research (cf.~\autoref{sec:stateOfTheArt}).
The resulting six quantitative privacy risk notifications provide in a transparent way realistic quantitative guarantees, \ie mathematically rigorous worst-case estimates (cf.~\autoref{sec:representingepsilon}), about the likelihood that certain sensitive data are revealed to unauthorized parties. 

Our study results provide first evidence that the inclusion of quantitative risk information does not significantly harm the objective understanding and, thus, the informed decision-making of the participants. Instead, the quantitative notifications perform similar to and, in some cases, slightly better (yet not significantly) than existing best practice by Xiong et al.~\cite{xiong_towards_2020}.
This is an important result, because it encourages further research into the use of
quantitative privacy risk notifications to transparently inform individuals about possible privacy risks when sharing their data. 

Our results also support the insights by Bullek et al.~\cite{bullek_towards_2017}, who highlight the usefulness of transparency when using DP mechanisms for understandability. Such understandability (cf~\cite{acquisti_privacy_2015, adjerid_sleights_2013}) allows for two potential modes for DP. Trusted authorities could recommend suitable values for $\varepsilon$ (global DP) and transparently communicate the resulting guarantees. Eventually, understanding of the guarantees would enable users to individually select~\cite{kohli_epsilon_2018} suitable values for $\varepsilon$ according to their privacy needs (local DP).

Concerning differences between the different risk communication formats, in our study we could not identify significant effects in general. Therefore, we cannot recommend any risk communication format over others. Having selected only the best practice formats from the medical domain, the difference between the formats might be too small to be detected. 
Additionally, in order to maintain comparability with the qualitative risk communication, we did not measure the understanding of the level of privacy protection, which should be communicated by the quantitative risk values. 
Future research will have to evaluate whether the additional quantitative privacy risk information contributes to a higher overall understanding of the involved privacy risk and which of the six quantitative risk communication formats perform best in this regard. 

\subsection{The Influence on Subjective and Objective Understanding}

The results regarding the overall effect of the notifications on subjective understanding show that the \emph{BaseLine} notification outperforms all quantitative notifications according to the mean value, and even significantly outperforms the condition \emph{FreqStatQuo}.

However, the results regarding objective understanding show a different behavior: no significant difference could be found, but two of the quantitative notifications, \ie \emph{FreqStatQuo} and \emph{PercPosNeg}, outperform the \emph{BaseLine} notification according to the mean value.
Consequently, we assume that all conditions had indeed similar effects on objective understanding.

However, the results regarding objective understanding were generally surprisingly low for all notifications: with four true/false questions the expected value of random guesses is two correct answers\footnote{Randomly guessing 4 questions with a probability of 50~\% each would result in an expected success rate of 2 answers}. The overall arithmetic mean on objective understanding in our sample set was $1.74$, with only the \emph{FreqPosNeg} notification achieving a result ($1.98$) close to this expected value (cf. \autoref{fig:SubjUnderstanding_avg}).
This suggests that the level of information provided even in the currently used notifications is insufficient for understanding the properties of DP. We discuss this aspect in more detail in future research (cf.~\autoref{sec:discussion_alternativeFormats}).

The two results on objective and subjective understanding seem to be contrary to each other. The quantitative notifications perform similarly to the \emph{BaseLine} notification on objective understanding, but less well on subjective understanding. 
This difference is especially pronounced regarding the notification \emph{FreqStatQuo}, which performs significantly lower than \emph{BaseLine} on subjective understanding, yet, achieves the highest overall value for objective understanding.
One possible explanation is that the \emph{BaseLine} condition creates a feeling of confidence, while the quantitative notification cause a more cautious reaction. 
However, this confidence caused by the \emph{BaseLine} notification ($5.6$ mean out of $7$) is not backed up by the objective understanding measured ($1.74$ mean out of $4$), as discussed above. Thus, we consider the high subjective understanding of the \emph{BaseLine} notification as overconfidence. 

Similar to research by Sirota et al.~\cite{sirota_1--x_2018}, future research should find privacy risk notifications, for which the subjective understanding corresponds best with the objective understanding.
Such a format would support informed consent, as it enables individuals to know when to seek more information, as opposed to \emph{dark patterns} (\eg~\cite{soe_circumvention_2020}), which try to maximize this overconfidence.
The framework presented by Bhatia et al.~\cite{bhatia_empirical_2018} could be a promising starting point for measuring the \emph{correct} perception of risk provided by the different privacy risk notifications.

\subsection{The Effect of Numeracy}
\label{sec:DiscussionNumeracy}

We discovered an effect of numeracy on subjective understanding significant for notification \emph{PercStatQuo}, and less pronounced for notifications \emph{FreqStatQuo}, \emph{PercPure} and \emph{BaseLine}.
The shape of the pattern (cf.~\autoref{fig:SubjUnderstanding_group_and_numeracy}) resembles the Dunning-Kruger effect~\cite{kruger_unskilled_2000}: low-skilled people also lack the meta-skill of accurately assessing their skill and tend to overestimate their skill level. Higher-skilled people, in contrast, can better assess their shortcomings and tend to underestimate their skill level. The same shape can be seen in our data.

According to Agrawal et al.~\cite{agrawal_exploring_2021} we already have a society, a technocracy, where the power to manage their privacy is reserved predominantly for people with experience in technology. The effect in our data only contributes further to this imbalance: individuals with the lowest and low numeracy are more confident of having understood the notification without grasping all the potentially severe consequences, leading to them sharing data more often and, therefore, being more vulnerable. As this effect is also found to a degree in the conventional \emph{BaseLine} notification this indicates a need for action.

With this knowledge, we see two different routes for counteracting the technocracy:
for some of the notifications we tested the effect was less prominent. Using these would give individuals a more accurate sense of their own understanding and enable them to seek more information before an important decision.
Alternatively, knowing this effect, privacy notifications could be tailored to the numeracy of individuals. 
Especially before more complex and important sharing decisions, the level of numeracy of an individual can be captured, for example, by a short questionnaire.
Individuals with low numeracy could then be cautioned while affirming individuals with higher numeracy, counteracting the confidence effect.
Personalization of risk communication has already been recommended in other contexts~\cite{tim_prior_risk_2015} and so called nudges have been shown to successfully influence individuals' privacy decisions.
However, a main concern with these approaches is that influence of a decision is inherently biased.
Our result provides an objective measure to evaluate the influence, \ie to counteract the existing effect, which disadvantages individuals with low numeracy.

\subsection{The Effect of Privacy Aptitude and Gender}
Concerning Privacy Aptitude, we found that participants with low privacy aptitude in the group \emph{FreqPure} performed less well in objective understanding. Since this is an isolated finding we cannot speculate about the reason for this performance. If this effect can be confirmed in future studies, the use of the frequencies format in privacy notification without additional information could not be recommended.

We could not find any other significant effects of Privacy Aptitude or Gender. However, the IUIPC test used in this study only records a self-reported privacy aptitude. Research about the accuracy of self-reported privacy aptitude has produced mixed results~\cite{keith_information_2013,wash_can_2017}. Using enhanced methods to record a more accurate privacy aptitude in future studies could yield additional significant results.


\section{Limitations and Future Work on Risk Communication Formats}
\label{sec:futureWork}

This study provides novel insights into using quantitative privacy risk notifications to inform laypeople about privacy risks when sharing their data.
However, our study has a number of limitations; at the same time, our results are hypothesis-generating and open up a number of promising research directions that may help to shape the way in which we enable people to share their data in a private manner. We next discuss both limitations and future research.

\subsection{Situating Privacy Risk Communication in the Real-World Context}

MTurk enabled us to quickly recruit a large sample within the budgetary requirements. However, we are aware of existing caveats caused by such a study design, especially regarding ethical and quality assurance concerns, and concerns regarding the representativeness of its users. 

To mitigate these concerns as much as possible, we considered existing academic (\eg~\cite{toxtli_quantifying_2021,ramirez_state_2021}) and practical\footnote{\eg Fair Crowd Work \url{http://faircrowd.work/platform/amazon-mechanical-turk}} recommendations for using MTurk. We informed, for example, all participants of any personal data collected before they accepted the task, provided means of aborting and deleting the collected data throughout the task, and handled submissions anonymously where possible. 

Secondly, using crowd workers as our participants, we have to consider MTurk's incentive structure: MTurk awards fixed payments per task. This encourages spending as little time as possible on reading the study texts and thinking through all possible repercussions of a decision~(cf.~\cite{downs_are_2010}).
We employed comprehension and attention checks to minimize the effect of this disadvantage and to ensure results comparable to conventional participation recruitment procedures (cf.~\cite{aguinis_mturk_2021}). 

Finally, we discussed the appropriateness of MTurk users for our research question. Recent surveys (\eg~\cite{thomas_validity_2017}) revealed that MTurk Workers are reasonably representative of the general population in the US (population validity). Thus we belief our results explore the design space of risk communication formats when employing DP in a representative way. However, transferring these insights into other cultural settings (\eg the European context) requires additional validation.

Furthermore, we acknowledge the participants' awareness of the study situation: 
Literature has shown, that privacy behavior is highly situated and responses concerned with hypothetical personal data might differ from actual personal data. In response we will follow up this initial exploration study with participatory design workshops (\eg~\cite{wong_eliciting_2017, bratteteig_cases_2014}) with citizens in the context of urban mobility. We will work closely with an institute for transport research that collects mobility data via smartphones to confirm and deepen the results in a realistic context. Such participatory design workshops will allow us to extend the variables considered, for example, how these descriptions relate to users' concrete sharing decisions, similar to work by Wu et al.~\cite{wu_effect_2012}. 

\subsection{Exploring Alternative Risk Communication Formats}
\label{sec:discussion_alternativeFormats}

Our findings show that quantitative privacy risk notifications perform comparably to a description that explains DP without quantitative information (\emph{BaseLine} condition derived from Xiong et al.~\cite{xiong_towards_2020}) in objective understanding. 
However, as noted before, the overall results in objective understanding were very low, which indicates general difficulties understanding the properties of DP.
We defined our quantitative privacy risk notifications with similar content and approximately the same length as the \emph{BaseLine} notification to ensure comparability. In future work, alternative privacy risk notifications (in terms of their content, length or their visual representation) might help to increase both the objective and subjective understanding of individuals and even avoid the possibly existing overconfidence effect. There has already been promising research done on visual aids in the field of medical risk communication and, concerning privacy, Xiong et al.~\cite{xiong_effect_2020} investigated the adaptation of a ``Fact Box''%
\footnote{A ``Fact Box'' is a visualization or tabular summary of data often used in a medical context to show dependencies (\eg of a particular disease and treatment) between one's position relative to others affected in the same situation (\eg \cite{brick_risk_2020}).} 
to explain DP. However, the results of their study are not yet available. 

We use our mathematically rigorous worst-case estimate regarding the likelihood that certain sensitive data are revealed to unauthorized parties. 
However, we limited our privacy risk notifications to one value for $\varepsilon$ only, as this is the default situation in current privacy decision settings. 
Future work could offer multiple values for $\varepsilon$ within the privacy decision setting. 
Bullek et al.~\cite{bullek_towards_2017} have already provided promising results for such a setting. They showed that explanations of DP can convey the relationship between the privacy protections of several DP options, however, their explanations have not been evaluated towards the absolute privacy risks of each option.
Explanations that can achieve both goals, \ie conveying the relation between different options and the absolute privacy protection of each option, would enable laypeople to select the most appropriate $\varepsilon$ value according to their personal privacy preferences in a specific situation.
Such a scenario could benefit from additional results in the medical domain, where the risks of different drugs often have to be compared to each other. 
We discovered in our literature review (cf.~\autoref{sec:communicatingrisk}) that certain risk communication formats are preferable in single risk communication, but others are recommended for comparing multiple risks. 
We, therefore, expect that the results of this study are not easily transferable to a setting when choosing between multiple $\varepsilon$ values. 
Future work is necessary for evaluating the benefits of such approaches.

Finally, our findings suggest that especially numeracy influence the effectiveness of privacy risk notifications in certain situations. A possible direction for future research might be to design tailored privacy risk notifications according to the individual's characteristics and, thus, equally support informed decision throughout the whole population. Such research can build on existing research in the area of explanations (\eg~\cite{miller_explanation_2019}). However, the need for better explanations has already been highlighted within an interview study by Agrawal et al.~\cite{agrawal_exploring_2021}, in which the participants suggested that explanations should be even part of a standardization process and included within software libraries. Such a standardization process is not restricted to the technical domain but might impact policy making as well, which we discuss next. 

\subsection{Towards Contextual Anonymity}

Our research is driven by the vision to enable individuals to make \emph{informed} decisions when sharing their data. An informed layperson should \emph{understand} the consequences of sharing personal data. Thus, we used results from medical risk communications together with a mathematically rigorous model of DP for the translation from the parameter $\varepsilon$ into one numeric privacy guarantee (cf.~\autoref{sec:representingepsilon}). The major advantage of this approach is that the privacy notification we displayed shows accurate, universally true bounds on the privacy risk rather than vague or averaged risk values, which disadvantage marginal groups. However, by applying the interpretative privacy risk model of Mehner et al.~\cite{mehner_towards_2021}, we rely on a worst-case estimate to derive a risk value. In more specific situations, where factors such as sample size or sensitivity of a query are known, other models might provide less conservative bounds, which might not sound as alarming as the worst-case estimates used here.
Future work should, in addition, consider such other models, for example, Naldi and D’Acquisto~\cite{naldi_differential_2015} or Hsu et al.~\cite{hsu_differential_2014}.

Furthermore, we hope that our research might provide the basis for making DP understandable and applicable to a broader range of people for making the abstract disadvantage of sharing the data, \ie the privacy risk, tangible. 
Current research focuses on showing which data is being exchanged for what purposes, for example, based on icons (\eg~\cite{cranor_informing_2021}), or visualizations (\eg~\cite{harkous_polisis_2018}). This research assumes that people are sharing their data, but what they are sharing is the risk of someone unauthorized accessing their personal information. We hope we can inform discussions in the policy domain by translating the abstract, intangible concept of \enquote{data sharing} into something more concrete. We envision the use of DP as a tool for \emph{contextual anonymity} that can be realized in two directions. 
Firstly, instead of communicating the purpose of data sharing, such as required by the General Data Protection Regulation, protection measures should be defined for specific areas of applications (\eg detailed location tracing versus on level of public transport stops). When applying DP, such protection measures define minimum requirements for the privacy budget $\varepsilon$ (cf.~\autoref{sec:defDP}). 
Secondly, based on such legally defined protection measures, laypeople can decide based on a privacy risk notification whether, in a concrete situation, the privacy protection proposed is sufficient for them and if not, lower the value of $\varepsilon$ to increase noise and, consequently, to increase their protection. 

\section{Conclusion}
\label{sec:conclusion}

Novel privacy-preserving technologies, like DP, protect the privacy of individuals, when used correctly. To support the sharing decisions, however, the privacy guarantees provided by these technologies need to be communicated to individuals in an understandable way (\eg~\cite{agrawal_exploring_2021}). 

Our research takes the first step in this direction by suggesting the use of risk communication formats informed by results from the medical domain. We combined these risk communication formats with a mathematically rigorous DP model, which translates the less intuitive privacy budget $\varepsilon$ into an understandable value for the privacy guarantee. With this combination, we achieve accurate quantitative privacy risk notifications, which not only inform about the use of DP, but also communicate the level of protection for the individual's privacy.

We evaluated the understandability of our proposed notifications in a crowd-sourced study against a well-performing conventional DP notification and found that our quantitative privacy risk notifications perform similarly in objective understanding. That makes the risk communication formats a promising research area. 
In subjective understanding, however, the quantitative privacy risk notifications lag behind the conventional qualitative notification. Future research will have to show whether additional risk communication tools, such as visual aids, can compensate this lack of confidence and increase the overall understandability.
Furthermore, we discovered that especially numeracy can influence the subjective understanding, including with the conventional DP communication.

In future studies we want to benefit people in two ways:
With a better understanding about supporting the sharing decision, using understandable and even tailored privacy risk notifications, we hope to enable individuals to accurately weigh the privacy risk against the benefit of sharing data. We hope that this encourages individuals, who are currently discouraged by the difficult to understand consequences, to share their data in a privacy-preserving way for legitimate causes.
Furthermore, we hope we can inform policy makers for a better \enquote{data security by design}. With improved communication on the privacy guarantees, we envision official minimal recommendations for the privacy budget $\varepsilon$ for different specific data collections contexts.


\bibliographystyle{ACM-Reference-Format}
\bibliography{references}

\appendix
\appendixpage

\section{Study resources}
\label{sec:studyResources}
\subsection{Welcome text}
\label{sec:introtext}
The aim of this survey is to evaluate the effectiveness of privacy notifications. It consists of up to 8 pages with 22 questions in total. We expect a working time of about 15 Minutes. It collects your response to a notification scenario, as well as some background information relevant to the analysis of your answers.
We collect no personal information except your gender. Since other studies identified influence of gender on the effectiveness of risk communication, we want to test if this is similar in our scenario. For this reason, we ask you to indicate your gender for a gender-specific analysis of the result.
Additionally, we automatically collect your IP address and the MTurk-IDs for quality assurance. After checking the data and approving the HIT towards Amazon Mechanical Turk, we will delete this data and your answer will be processed anonymously. All data collected from you will be used exclusively for study purposes and will not be passed on to third parties.
We record your response during your participation. If you change your mind and don't want to continue to participate in the study, you can delete your answers by clicking the Button "Exit and clear survey".
Please do not use the browser navigation (for example the "back" button) during the survey.
\textbf{Please keep in mind that this survey aims to evaluate the effectiveness of the notification, not your performance or knowledge.}

\subsection{Scenario Description}
\label{sec:scenarioDescription}
You are a long-time customer of the ridesharing service "\textbf{CityCar}" in your home town.
Recently, the CityCar app was updated.
Since this update, you have not booked a ride, but today, you feel like booking a ride again – it is snowy, and public transport availability is limited.
After opening the app, you see a \textbf{notification}. Such notification is unfamiliar to you, before, the app had never shown such a notification.
The notification in the CityCar app informs you that aggregated statistics about the \textbf{visited locations} of all customers are \textbf{transmitted} to the Urban Redevelopment Authority. Such aggregated statistics include, for example, the "number of customers in the last month", which are reported for each location to the Urban Redevelopment Authority.
The notification also explains that the customer data will be used, among other things, to \textbf{improve the urban infrastructure} and public transport, especially in areas with high utilization of ridesharing services.
Together with this notification, you receive a short explanation of the potential \textbf{privacy risk} of you being re-identified in the aggregated statistics ...

\begin{figure}[tb]
    \centering
    \includegraphics[width=0.6\columnwidth]{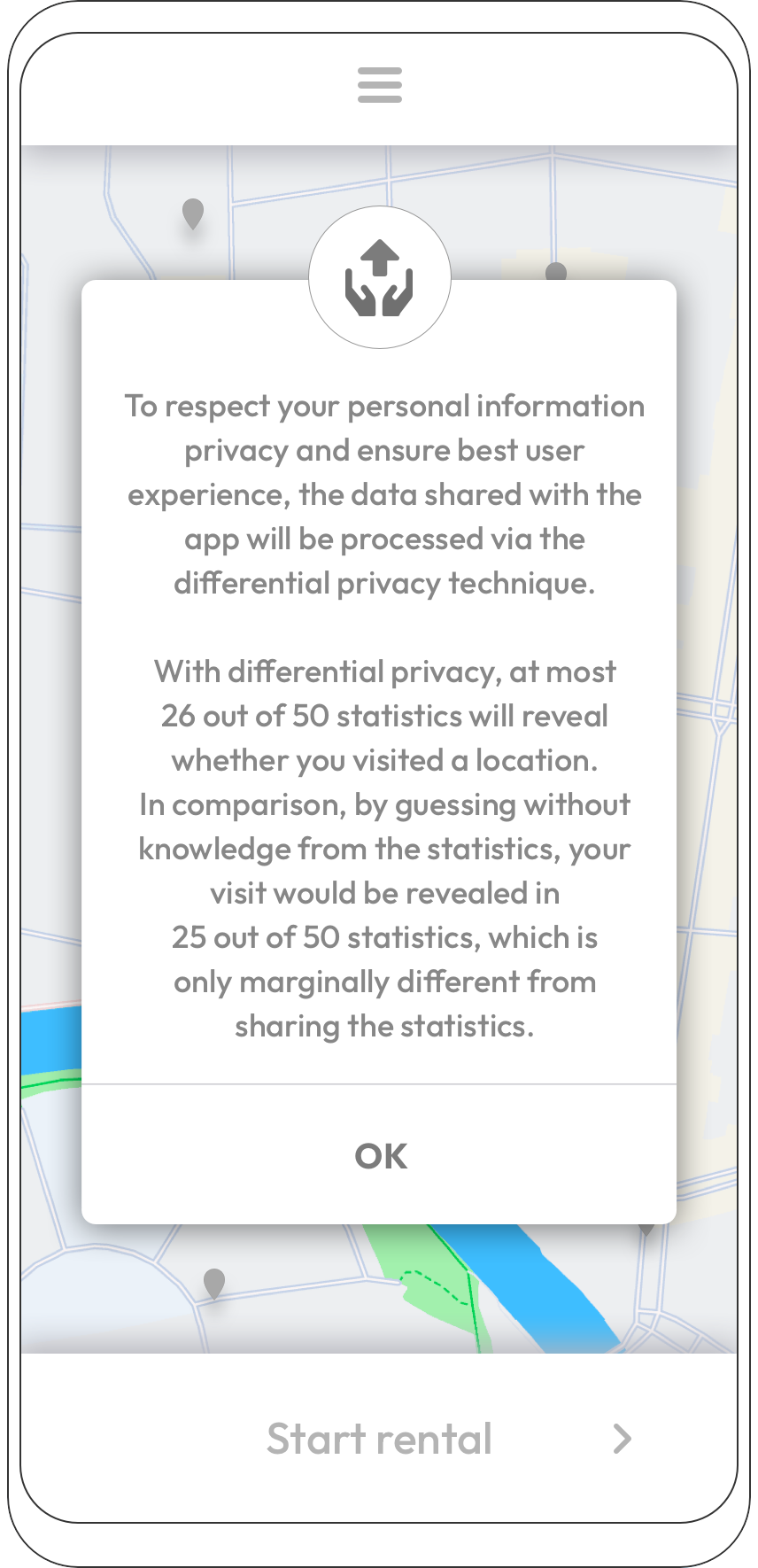}
    \Description[Example of the notification shown to the participants]{Mock of a mobile phone showing a street map and a button \enquote{Start rental} in the background. The foreground displays a pop-up containing the text of the FreqStatQuo condition and a button \enquote{OK}.}
    \caption{Visual Representation: \emph{FreqStatQuo}}
    \label{fig:condition}
\end{figure}

\subsubsection{Scenario-attention check questions}
\setlist[itemize]{leftmargin=3mm}
\begin{itemize}
    \item What kind of app is mentioned in the scenario? [Banking app, \underline{Ridesharing app}, Messaging app]
    \item What kind of data is transmitted to the Urban Redevelopment Authority? [\underline{Location data}, Usage statistics, Preferred car types]
    \item What is indicated in the notification?[Guarantee, \underline{Privacy risk}, Transparency]
\end{itemize}

\subsection{Understanding questions}
\label{sec:UnderstandingQuestions}
\subsubsection{Subjective Understanding}
\begin{itemize}
    \item I have understood the notification.
    \item After reading the notification, I can assess my privacy risk, that is, the risk that the Urban Redevelopment Authority knows whether I visited a location or not after seeing the aggregated statistics. 
    \item The notification provided all the information I wanted to know about the used technique (differential privacy).
\end{itemize}

\subsubsection{Objective Understanding}
\begin{itemize}
    \item In the scenario, the Urban Redevelopment Authority is less/ more likely to know whether I visited the city hospital than if CityCar provided the unprotected data. (less)
    \item In the scenario, my risk of the Urban Redevelopment Authority knowing whether I visited the city hospital is higher/lower than if the CityCar app does not share any data. (higher)
    \item Without access to CityCar's data, there is no risk that the Urban Redevelopment Authority will know whether I visited the city hospital. (false)
    \item In the scenario, the Urban Redevelopment Authority can determine whether I visited the city hospital or not. (false)
\end{itemize}

\subsection{Privacy Risk Notifications}

\begin{figure*}[tb]
    \centering
    \small
    \begin{tabularx}{\textwidth}{lX}\toprule
        Condition   & Notification wording\\\midrule
FreqPure    & To respect your personal information privacy and ensure best user experience, the data shared with the app will be processed via the differential privacy technique.With differential privacy, at most 26 out of 50 statistics will reveal whether you visited a location. \vspace{0.7em}\\
FreqPosNeg  & To respect your personal information privacy and ensure best user experience, the data shared with the app will be processed via the differential privacy technique.With differential privacy, at most 26 out of 50 statistics will reveal whether you visited a location. For 24 out of 50 statistics, however, your visit remains undetected. \vspace{0.7em}\\
FreqStatQuo & To respect your personal information privacy and ensure best user experience, the data shared with the app will be processed via the differential privacy technique. With differential privacy, at most 26 out of 50 statistics will reveal whether you visited a location. In comparison, by guessing without knowledge from the statistics, your visit would be revealed in 25 out of 50 statistics, which is only marginally different from the case with the statistics.   \vspace{0.7em}\\
PercPure    & To respect your personal information privacy and ensure best user experience, the data shared with the app will be processed via the differential privacy technique. With differential privacy, at most 52\% of the statistics will reveal whether you visited a location. \vspace{0.7em}\\
PercPosNeg  & To respect your personal information privacy and ensure best user experience, the data shared with the app will be processed via the differential privacy technique.With differential privacy, at most 52\% of the statistics will reveal whether you visited a location. In 48\% of the statistics, however, your visit remains undetected.\vspace{0.7em}\\
PercStatQuo & To respect your personal information privacy and ensure best user experience, the data shared with the app will be processed via the differential privacy technique. With differential privacy, at most 52\% of the statistics will reveal whether you visited a location. In comparison, by guessing without knowledge from the statistics, your visit would be revealed in 50\% of the statistics, which is only marginally different from the case with the statistics.\vspace{0.7em}\\
BaseLine    & To respect your personal information privacy and ensure best user experience, the data shared with the app will be processed via the differential privacy technique. That is, the app company will store your data but only use the aggregated statistics with modification so that your personal information cannot be learned. However, your personal information may be leaked if the company’s database is compromised. \\
        \bottomrule
    \end{tabularx}
    \caption{Notification Conditions}
    \label{tab:conditions}       
\end{figure*}
\label{sec:ConditionsText}
The wording of the 7 notifications is presented in \autoref{tab:conditions} and an example of the visual representation is shown in \autoref{fig:condition}

\section{Statistic resources}
The details results from the linear regression models are as follows:
\subsection{Subjective Understanding}
\label{sec:statistic_resources_subjective}

\begin{tiny}
\begin{verbatim}
    Residuals:
    Min      1Q  Median      3Q     Max 
-3.9382 -0.6814  0.1885  0.7612  2.6016 

Coefficients:
                         Estimate Std. Error t value Pr(>|t|)    
(Intercept)               5.74347    0.55034  10.436   <2e-16 ***
GruppeG1                 -0.60041    0.85960  -0.698   0.4854    
GruppeG2                 -0.34916    0.73273  -0.477   0.6341    
GruppeG3                 -1.37402    0.71754  -1.915   0.0565 .  
GruppeG4                 -0.96664    0.69193  -1.397   0.1635    
GruppeG5                 -0.30527    0.74248  -0.411   0.6813    
GruppeG6                 -1.56866    0.69863  -2.245   0.0255 *  
Numeracylowest           -0.05163    0.58916  -0.088   0.9302    
Numeracylow              -0.10838    0.69311  -0.156   0.8758    
Numeracyhigh             -0.65603    0.70904  -0.925   0.3556    
P_binnedniedrig          -0.02355    0.71636  -0.033   0.9738    
P_binnedhoch             -0.22716    0.40733  -0.558   0.5775    
Genderfemale              0.33008    0.41217   0.801   0.4239    
GruppeG1:Numeracylowest   0.94925    0.78833   1.204   0.2295    
GruppeG2:Numeracylowest   0.60341    0.73303   0.823   0.4111    
GruppeG3:Numeracylowest   0.67367    0.74873   0.900   0.3690    
GruppeG4:Numeracylowest   0.92976    0.74979   1.240   0.2160    
GruppeG5:Numeracylowest   0.21832    0.74455   0.293   0.7696    
GruppeG6:Numeracylowest   1.37512    0.76692   1.793   0.0740 .  
GruppeG1:Numeracylow      0.63758    1.02592   0.621   0.5348    
GruppeG2:Numeracylow      0.42610    0.92291   0.462   0.6447    
GruppeG3:Numeracylow      0.84179    0.94395   0.892   0.3732    
GruppeG4:Numeracylow      1.57894    0.93063   1.697   0.0908 .  
GruppeG5:Numeracylow     -0.28900    0.93189  -0.310   0.7567    
GruppeG6:Numeracylow      2.14711    0.95159   2.256   0.0248 *  
GruppeG1:Numeracyhigh    -0.25199    1.00130  -0.252   0.8015    
GruppeG2:Numeracyhigh     0.58184    0.98730   0.589   0.5561    
GruppeG3:Numeracyhigh    -0.36557    0.96798  -0.378   0.7060    
GruppeG4:Numeracyhigh    -0.05491    1.02103  -0.054   0.9572    
GruppeG5:Numeracyhigh     0.97992    1.05694   0.927   0.3546    
GruppeG6:Numeracyhigh     0.48464    0.97167   0.499   0.6183    
GruppeG1:P_binnedniedrig  0.01333    0.95403   0.014   0.9889    
GruppeG2:P_binnedniedrig -0.85984    0.83482  -1.030   0.3039    
GruppeG3:P_binnedniedrig  0.62718    0.91725   0.684   0.4947    
GruppeG4:P_binnedniedrig -0.36457    1.07912  -0.338   0.7357    
GruppeG5:P_binnedniedrig -0.25461    0.83713  -0.304   0.7612    
GruppeG6:P_binnedniedrig -0.76502    0.87354  -0.876   0.3819    
GruppeG1:P_binnedhoch    -0.50195    0.65089  -0.771   0.4412    
GruppeG2:P_binnedhoch    -1.08426    0.57889  -1.873   0.0621 .  
GruppeG3:P_binnedhoch    -0.69930    0.58525  -1.195   0.2331    
GruppeG4:P_binnedhoch    -0.21058    0.59117  -0.356   0.7219    
GruppeG5:P_binnedhoch    -0.39231    0.61283  -0.640   0.5226    
GruppeG6:P_binnedhoch     0.40118    0.61525   0.652   0.5149    
GruppeG1:Genderfemale    -0.97440    0.60732  -1.604   0.1097    
GruppeG2:Genderfemale    -0.25795    0.54579  -0.473   0.6368    
GruppeG3:Genderfemale     0.48597    0.57036   0.852   0.3949    
GruppeG4:Genderfemale    -0.58231    0.59244  -0.983   0.3265    
GruppeG5:Genderfemale    -0.60255    0.56762  -1.062   0.2893    
GruppeG6:Genderfemale    -0.07864    0.60864  -0.129   0.8973    
---
Signif. codes:  0 ‘***’ 0.001 ‘**’ 0.01 ‘*’ 0.05 ‘.’ 0.1 ‘ ’ 1

Residual standard error: 1.264 on 291 degrees of freedom
Multiple R-squared:  0.2736,	Adjusted R-squared:  0.1537 
F-statistic: 2.283 on 48 and 291 DF,  p-value: 1.577e-05
\end{verbatim}
\end{tiny}

\subsection{Objective Understanding}
\label{sec:statistic_resources_objective}
\begin{tiny}
\begin{verbatim}
    Residuals:
     Min       1Q   Median       3Q      Max 
-2.40037 -0.63293  0.00534  0.65711  2.35053 

Coefficients:
                          Estimate Std. Error t value Pr(>|t|)    
(Intercept)               1.613962   0.413580   3.902 0.000118 ***
GruppeG1                  0.241608   0.645981   0.374 0.708664    
GruppeG2                 -0.025616   0.550644  -0.047 0.962928    
GruppeG3                  0.280895   0.539228   0.521 0.602818    
GruppeG4                  0.428400   0.519981   0.824 0.410685    
GruppeG5                  0.036928   0.557970   0.066 0.947278    
GruppeG6                 -0.071511   0.525019  -0.136 0.891752    
Numeracylowest            0.077447   0.442752   0.175 0.861263    
Numeracylow              -0.163815   0.520869  -0.315 0.753364    
Numeracyhigh              0.171542   0.532841   0.322 0.747731    
P_binnedniedrig           0.535320   0.538342   0.994 0.320861    
P_binnedhoch              0.111571   0.306103   0.364 0.715758    
Genderfemale              0.046542   0.309742   0.150 0.880664    
GruppeG1:Numeracylowest   0.204641   0.592425   0.345 0.730021    
GruppeG2:Numeracylowest  -0.501209   0.550867  -0.910 0.363652    
GruppeG3:Numeracylowest  -0.084992   0.562670  -0.151 0.880040    
GruppeG4:Numeracylowest  -0.937851   0.563465  -1.664 0.097102 .  
GruppeG5:Numeracylowest  -0.286620   0.559528  -0.512 0.608863    
GruppeG6:Numeracylowest   0.065419   0.576340   0.114 0.909706    
GruppeG1:Numeracylow      0.330996   0.770971   0.429 0.668005    
GruppeG2:Numeracylow      0.022382   0.693562   0.032 0.974278    
GruppeG3:Numeracylow      0.286116   0.709373   0.403 0.686997    
GruppeG4:Numeracylow     -0.885753   0.699360  -1.267 0.206340    
GruppeG5:Numeracylow      0.148045   0.700311   0.211 0.832724    
GruppeG6:Numeracylow      0.187067   0.715112   0.262 0.793822    
GruppeG1:Numeracyhigh     0.594622   0.752471   0.790 0.430039    
GruppeG2:Numeracyhigh    -0.897780   0.741947  -1.210 0.227248    
GruppeG3:Numeracyhigh     0.450249   0.727434   0.619 0.536430    
GruppeG4:Numeracyhigh    -0.127134   0.767302  -0.166 0.868516    
GruppeG5:Numeracyhigh    -1.195565   0.794288  -1.505 0.133356    
GruppeG6:Numeracyhigh    -0.265667   0.730201  -0.364 0.716252    
GruppeG1:P_binnedniedrig -1.532084   0.716946  -2.137 0.033435 *  
GruppeG2:P_binnedniedrig -0.210960   0.627365  -0.336 0.736915    
GruppeG3:P_binnedniedrig -0.412925   0.689310  -0.599 0.549611    
GruppeG4:P_binnedniedrig -0.639017   0.810956  -0.788 0.431350    
GruppeG5:P_binnedniedrig -0.108807   0.629100  -0.173 0.862806    
GruppeG6:P_binnedniedrig -0.571165   0.656459  -0.870 0.384979    
GruppeG1:P_binnedhoch    -0.332936   0.489140  -0.681 0.496631    
GruppeG2:P_binnedhoch     0.356772   0.435031   0.820 0.412826    
GruppeG3:P_binnedhoch     0.007763   0.439812   0.018 0.985930    
GruppeG4:P_binnedhoch     0.266738   0.444259   0.600 0.548701    
GruppeG5:P_binnedhoch     0.086576   0.460542   0.188 0.851018    
GruppeG6:P_binnedhoch    -0.469945   0.462353  -1.016 0.310274    
GruppeG1:Genderfemale    -0.494356   0.456398  -1.083 0.279630    
GruppeG2:Genderfemale     0.239656   0.410158   0.584 0.559471    
GruppeG3:Genderfemale    -0.278547   0.428626  -0.650 0.516294    
GruppeG4:Genderfemale    -0.092387   0.445216  -0.208 0.835756    
GruppeG5:Genderfemale     0.576082   0.426563   1.351 0.177898    
GruppeG6:Genderfemale     0.204645   0.457393   0.447 0.654907    
---
Signif. codes:  0 ‘***’ 0.001 ‘**’ 0.01 ‘*’ 0.05 ‘.’ 0.1 ‘ ’ 1

Residual standard error: 0.9499 on 291 degrees of freedom
Multiple R-squared:  0.1655,	Adjusted R-squared:  0.02785 
F-statistic: 1.202 on 48 and 291 DF,  p-value: 0.1825
\end{verbatim}
\end{tiny}

\subsection{Additional Results}
Throughout the statistical analysis we considered additional statistical models, the results of which were not deemed essential in light of the later results presented in this paper. For the sake of completeness and accountability, we want to share these additional models here.

\subsubsection{ANOVA model of Individuals' Characteristics}
Before calculating the linear regression model in \autoref{sec:results_covariates}, we used another ANOVA model to get an impression of the effects of these characteristics. This ANOVA model included 5 predicates as IVs (condition, numeracy, privacy aptitude, interaction of condition with numeracy, interaction of condition with privacy aptitude) and either the subjective or the objective as DV.
Due to the categorical values of the gender covariate, we did not include gender in this model.

In the model for the subjective understanding, we find one significant effect for the numeracy ($p=0.0007152$). The correlation is negative, indicating that the subjective understanding decreases as numeracy increases. This result is in line with the results found in the linear regression model and discussed in detail in the main paper. Since the linear regression shows the effect in more detail, we omitted this model from the main paper.

The equivalent ANOVA model for the objective understanding did not indicate any significant effects ($p\ge 0.1093$) similar to the equivalent linear regression model covered in \autoref{sec:results_covariates}.

\subsubsection{Outlier analysis}
Even though we decided against outlier removal in the main analysis of this paper, we investigated into potential outliers due to value differences. 

We identified 14 outlier in the data set, by allowing up to 1.5 times of the interquartile range outside the lower or upper quartile~\footnote{https://statistikguru.de/spss/einfaktorielle-anova/voraussetzungen-5.html}.
The values of these ourliers are shown in \autoref{fig:outlier_subj}.

\begin{table}[tb]
    \centering
    \scriptsize
        \begin{tabularx}{\columnwidth}{@{}llllll@{}}
            \toprule
            \begin{tabular}[c]{@{}l@{}}Subj. \\Underst.\end{tabular} &
                 \begin{tabular}[c]{@{}l@{}}Obj.\\ Underst.\end{tabular} & 
                 Group & Num. & 
                 \begin{tabular}[c]{@{}l@{}}Priv.\\Apt.\end{tabular}& 
                 Gender\\\midrule
            1.000000 & 2 & G2 & 4 & 6.750000 & male \\
            1.000000 & 2 & G1 & 4 & 6.333333 & female\\
            2.333333 & 3 & G5 & 4 & 6.833333 & male\\
            2.000000 & 3 & G4 & 4 & 5.833333 & male\\
            1.666667 & 3 & G2 & 4 & 6.750000 & female\\
            2.000000 & 1 & G4 & 4 & 4.833333 & male\\
            1.000000 & 3 & G2 & 1 & 7.000000 & male\\
            2.666667 & 2 & G5 & 2 & 4.875000 & female\\
            2.000000 & 2 & G4 & 3 & 6.750000 & male\\
            1.000000 & 2 & G1 & 3 & 6.875000 & male\\
            2.000000 & 0 & G4 & 1 & 5.666667 & female\\
            1.666667 & 1 & G5 & 1 & NA & male\\
            2.666667 & 3 & G5 & 4 & 5.000000 & female\\
            2.333333 & 2 & G4 & 4 & 7.000000 & male\\\bottomrule
        \end{tabularx}
    \caption{Values of Potential Outliers}
    \label{fig:outlier_subj}
\end{table}

In the data set without the outliers, a correlation between the conditions and the results from the privacy aptitude test could not be rejected (ANOVA with $p=0.0184$). With the full data set, the mentioned correlation was rejected~(ANOVA with $p=0.2563$) and, therefore, the result presented in the main part of the paper are sound. Furthermore, one can see that a majority of the outliers achieved the highest numeracy score of 4, which indicates that the questions were considered carefully. Therefore, in accordance with the main paper the submissions should not be regarded as outliers for the analysis of the investigated effects. 

In order to investigate possible causes for this correlation, we performed a post-hoc t-test using the privacy aptitude as the output variable, resulting in no significant effects of one single condition over another~(cf. \autoref{fig:pairwisePrivApt}). 
An effect of the conditions on the results of the privacy aptitude test can, therefore, not be confirmed.

\begin{table}
    \centering
    \scriptsize
    \begin{tabularx}{\columnwidth}{@{}lllllll@{}}\toprule
                              & (1)  & (2)  & (3)  & (4)  & (5)  & (6)  \\ \midrule
        FreqPure \hfill(1)    & -    & -    & -    & -    & -    & -    \\
        FreqPosNeg \hfill(2)  & 1.00 & -    & -    & -    & -    & -    \\
        FreqStatQuo \hfill(3) & 1.00 & 1.00 & -    & -    & -    & -    \\
        PercPure \hfill(4)    & 1.00 & 1.00 & 1.00 & -    & -    & -    \\
        PercPosNeg \hfill(5)  & 1.00 & 1.00 & 1.00 & 1.00 & -    & -    \\
        PercStatQuo \hfill(6) & 0.21 & 0.48 & 1.00 & 0.61 & 0.63 & -    \\
        BaseLine \hfill(7)    & 0.78 & 1.00 & 1.00 & 1.00 & 1.00 & 1.00 \\\bottomrule
    \end{tabularx}
    \caption{Pairwise t-test (Privacy aptitude)}
    \label{fig:pairwisePrivApt}
\end{table}

Nevertheless, we want to discuss possible causes of the results of the first ANOVA model.
Since we assigned participants randomly to each condition, there should not be an effect of the group on the privacy aptitude by design. We provide two possible reasons for this relationship. 

Firstly, throughout our study, we provided participants with the option to quit and delete all answers recorded at any point in the study. Different levels of privacy aptitude might prompt participants to use this option in different circumstances: Participants with high privacy aptitude, for example, might have quit more often when seeing one condition and participants with lower privacy aptitude might have quit more often when seeing another condition. Unfortunately we cannot confirm this because we removed all records of the aborted study sessions.
Secondly, another possible explanation for the relationship is the order of questions in our questionnaire: we did not randomize the order of the parts of the survey for technical and operational reasons. All participants answered the privacy aptitude questions after reading the conditions. 
Due to this order, it is possible that the conditions temporarily influenced the participants to answer differently in our privacy aptitude test. 

This observation might relate to the well-known \emph{privacy paradox}~\cite{acquisti_privacy_2005}, which describes the difference between the privacy behaviour of individuals and their privacy aptitude measured separately. In our study, having read the privacy notifications, the participants answers in the \emph{Internet Users' Information Privacy Concerns} questionnaire might not reflect their actual privacy aptitude, but instead were influenced by the condition. In the worst-case, this effect could further increase the privacy paradox, when notifications intentionally or unintentionally manipulate the participants behaviour further away from their actual privacy aptitude.
Alternatively, this effect could be used for the benefit of the individual. The effect could be reduced by choosing notification formats which act in opposition to the privacy paradox.
Similar work has been done under the name of \emph{nudges} (\eg~\cite{diaz_ferreyra_preventative_2020}).
Since we could neither reject nor confirm the effect between the condition and the privacy aptitude in our study, further research might shed more light on this relationship.

\section{Risk communication Formats}
\label{sec:riskFormatsDetails}
\autoref{tab.formats} summarizes all discovered medical risk communication formats.
\newcolumntype{L}{>{\raggedright\arraybackslash}X}
\begin{table}
\centering
\begin{sideways}
\scriptsize
\begin{tabularx}{\dimexpr\textheight-4\baselineskip\relax}{@{}>{\hsize=0.4\hsize}L>{\hsize=0.6\hsize}L>{\hsize=0.6\hsize}L>{\hsize=0.6\hsize}L>{\hsize=1\hsize}L@{}}
\toprule
Format & Description & Examples & Possible Variations & Recommendations / Comments \\ \toprule
Percentage\newline\cite{trevena_presenting_2013, ancker_covid-19_2020, bonner_current_2021} & Relative frequency with baseline 100 & \textless{}1\% of the participants are re-identified&
\vspace{-1em}\begin{itemize}[leftmargin=0.5em]
    \item Describe pos. and neg. outcomes, 
    \item Compare to Known Risk / Peers / status quo
\end{itemize} & 
\vspace{-1em}\begin{itemize}[leftmargin=0.5em]
    \item Recommended for comparing multiple risks
    \item Avoid decimal point $\to$ rounding
    \item $0\%>p>1\%$ as $<1\%$
    \item Percentages better for comparing two events than frequencies
\end{itemize}  \\ \midrule 
Simple Frequencies \newline\cite{gigerenzer_helping_2007, mcdowell_simple_2016, trevena_presenting_2013, bonner_current_2021,brick_risk_2020} &  Present the number of affected people in relation to meaningful baseline  & 3 out of 350 participants are re-identified &
\vspace{-1em}\begin{itemize}[leftmargin=0.5em]
    \item Describe pos. and neg. outcomes
    \item Compare to known Risk / Peers / status quo
\end{itemize} &  
\vspace{-1em}\begin{itemize}[leftmargin=0.5em]
    \item Denominators
    \begin{itemize}
        \item consistent
        \item as small as possible
        \item powers of 10
    \end{itemize}
    \item avoid 1 as numerator
    \item specify time frame
\end{itemize}\\ \midrule 
Fractions\newline\cite{bansback_communicating_2017} & Present the fraction of affected people as textual representation & Half of all statistics will reveal whether you visited a location & Variations not feasible due to accuracy &
\vspace{-1em}\begin{itemize}[leftmargin=0.5em]
    \item Well understood, if representation is available
\end{itemize} \\ \midrule 
Numbers needed to treat \newline\cite{gigerenzer_helping_2007, trevena_presenting_2013,spiegelhalter_risk_2017,visschers_probability_2009} & How many people need to do the action, before 1 is expected to be affected by the risk & 117 Participants need to participate before we expect a participant to be re-identified & 
\vspace{-1em}\begin{itemize}[leftmargin=0.5em]
    \item Compare to known Risk / Peers / status quo
\end{itemize} &  
\vspace{-1em}\begin{itemize}[leftmargin=0.5em]
    \item discouraged by \cite{spiegelhalter_risk_2017,visschers_probability_2009}
\end{itemize}\\ \midrule 
1-in-x\newline\cite{bonner_current_2021,pighin_1--x_2011,sirota_1--x_2018} & Present 1 in how many people are effected & about 1 in 117 &
\vspace{-1em}\begin{itemize}[leftmargin=0.5em]
    \item Compare to known Risk / Peers / status quo 
\end{itemize}\vspace{-1em}\null &  
\vspace{-1em}
\begin{itemize}[leftmargin=0.5em]
    \item hard to understand 
    \item wrong intuition
    \item widely discouraged 
\end{itemize}\vspace{-1em}\null\\ \midrule 
Odds\newline\cite{visschers_probability_2009} & $\frac{\nicefrac{x_Y}{x_N}}{\nicefrac{x_Y}{y_Y}}$ \newline with $x_Y$: revealed people participating,\newline $x_N$: not revealed people participating\newline $y_Y$: people revealed not participating\newline $y_N$: people not revealed, not participating & The odds of being revealed by a statistics are 1.083. &
always with: comparison to status quo
\begin{itemize}[leftmargin=0.5em]
    \item Compare to known Risk / Peers
\end{itemize} &
\begin{itemize}[leftmargin=0.5em]
    \item Not well understood for representation of risks
\end{itemize}\\\bottomrule
\end{tabularx}
\end{sideways}
\caption{Risk Communication Formats in Medical Research}
\label{tab.formats}
\end{table}
\end{document}